\documentclass[12pt]{article}

\usepackage{graphicx}
\usepackage{amssymb}
\usepackage{amsmath}
\usepackage{booktabs}
\usepackage{colortbl}

\newcommand{\tinymath}[1]{{{\mbox{\tiny $#1$}}}}
\newcommand{\zp}{Z^{\prime}}
\newcommand{\mzp}{M_{\tilde{Z}^\prime}}
\newcommand{\uonep}{\ensuremath{U(1)^\prime}}
\newcommand{\gz}{\ensuremath{g_{\tinymath{Z^\prime}}}}

\newcommand{\refeq}[1]{Eq.~\ref{#1}}
\newcommand{\ctoprule}{\toprule[0.5mm]}
\newcommand{\cbottomrule}{\bottomrule[0.5mm]}
\newcommand{\graycell}{\cellcolor[rgb]{0.8,0.8,0.8}}

\begin{document}

\begin{titlepage}

\begin{flushright}
UG-FT-261/09\\
CAFPE-131/09\\
EFI 09-28\\
November 10, 2009
\end{flushright}

\vspace{0.7cm}
\begin{center}
\Large\bf\boldmath Combining Anomaly and $\zp$ Mediation
of Supersymmetry Breaking  \unboldmath
\end{center}

\vspace{0.8cm}
\begin{center}
{\sc Jorge de Blas$^{(a)}$, Paul Langacker$^{(b)}$, Gil Paz$^{(c)}$, Lian-Tao Wang$^{(d)}$} \\
\vspace{0.7cm}
{\sl ${}^a$ Departamento de F{\'\i}sica Te\'orica y del Cosmos and CAFPE, \\
Universidad de Granada, E-18071 Granada, Spain \\[0.3cm]
${}^b$\, School of Natural Sciences, Institute for Advanced Study\\  Princeton, NJ 08540, U.S.A\\[0.3cm]
${}^c$\, Enrico Fermi Institute, University of Chicago\\ 5640 S. Ellis Ave., Chicago, IL 60637, U.S.A\\[0.3cm]
${}^d$\, Department of Physics, Princeton University \\Princeton, NJ 08544, U.S.A }
\end{center}

\vspace{1.0cm}
\begin{abstract}
\vspace{0.2cm} \noindent 
We propose a scenario in which the supersymmetry breaking effect
mediated by an additional $U(1)'$ is comparable with that of anomaly
mediation. We argue that such a scenario can be naturally realized in
a large class of models.  Combining anomaly with $\zp$ mediation
allows us to solve the tachyonic slepton problem of the former and
avoid significant fine tuning in the latter. We focus on an NMSSM-like
scenario where $U(1)'$ gauge invariance is used to forbid a tree-level
$\mu$ term, and present concrete models, which admit successful
dynamical electroweak symmetry breaking.  Gaugino masses are somewhat
lighter than the scalar masses, and the third generation squarks are
lighter than the first two. In the specific class of models under
consideration, the gluino is light since it only receives a
contribution from 2-loop anomaly mediation, and it decays dominantly
into third generation quarks. Gluino production leads to distinct LHC
signals and prospects of early discovery. In addition, there is a
relatively light $\zp$, with mass in the range of several
TeV. Discovering and studying its properties can reveal important
clues about the underlying model.
\end{abstract}
\vfil

\end{titlepage}

\section{Introduction}

Many top-down supersymmetric constructions contain extra abelian gauge
interactions \cite{zp-rev}. Since a $U(1)'$ often couples to both the
minimal supersymmetric standard model (MSSM) and additional hidden
sectors, it is plausible that it plays a role in mediating
supersymmetry breaking.  One might refer to such a scenario as
$Z^\prime$ mediation.  There are several possibilities for an extra
$Z'$ to participate in the mediation of supersymmetry breaking.
Completely analogous to the vector supermultiplet of the Standard
Model (SM) gauge groups, a $U(1)'$ can be the mediator of supersymmetry
breaking either through gauge mediation \cite{gaugemed} or gaugino
mediation \cite{Kaplan:1999ac,Chacko:1999mi} . In this paper, we will
use $Z'$ mediation to collectively refer to both possibilities, and
give specific qualification when referring to particular realizations.

In a pair of previous works \cite{Langacker:2007ac,Langacker:2008ip},
we have considered a special implementation of such a scenario, in
which the $Z^\prime$ gaugino becomes massive as a result of
supersymmetry breaking.  Although it was referred to as ``$Z^\prime$
mediation'' in \cite{Langacker:2007ac,Langacker:2008ip}, it can be
thought of as a $Z'$-gaugino mediated supersymmetry breaking, and this
is the name we will use in this paper\footnote{ The mechanism under
which the $Z^\prime$ gaugino becomes massive was left unspecified in
\cite{Langacker:2007ac,Langacker:2008ip}. Although we refer to this
situation as $Z'$-gaugino mediation, the name should \emph{not}
necessarily imply an underlying extra dimensional model, as in the
original gaugino mediation papers
\cite{Kaplan:1999ac,Chacko:1999mi}.}.  A possible realization of this
scenario in string theory was subsequently proposed in
\cite{Verlinde:2007qk}. Further applications and realizations of the
scenario are discussed in \cite{Kikuchi:2008xu} and
\cite{Grimm:2008ed}, respectively.  Extensions of the Higgs sector of
such a scenario have been discussed in \cite{Langacker:2008dq}. A
generic feature of the $\zp$-gaugino mediation scenario is the
generation of the soft scalar masses at one-loop order and gaugino
masses at two-loop order
\cite{Langacker:2007ac,Langacker:2008ip}. Simple estimates implies
that the scalar masses are about 1000 times heavier than the MSSM
gaugino masses.  Since direct searches constrain the gaugino masses to
be above $\sim 100$ GeV, it follows that if the MSSM gaugino masses
are generated by $\zp$-gaugino mediation and are in the range of
100-1000 GeV, the soft scalar masses are in the range 100-1000 TeV. To
obtain electroweak symmetry breaking at its observed scale, one fine
tuning is needed.

In this article, we study the possibility that the effect of $Z'$
mediation can be comparable with some other supersymmetry breaking
mediation mechanism. By choosing flavor universal $U(1)'$ charges, the
$Z'$ mediation is naturally flavor diagonal. Hence, we would like to
narrow our attention to mechanisms with similar properties in order to
avoid introducing additional tuning or new flavor protection
mechanisms. Typical examples of such mechanisms are gauge mediation
\cite{gaugemed}, anomaly mediation
\cite{Randall:1998uk,Giudice:1998xp}, and gaugino mediation
\cite{Kaplan:1999ac,Chacko:1999mi}. Combining $Z'$ mediation with
gauge mediation or gaugino mediation amounts to straightforward
extensions of these scenarios with a larger gauge symmetry. These
scenarios are of phenomenological interest, but we will not pursue
them further in this paper. Instead, we focus on the possibility of a
$Z'$-gaugino mediation that is co-dominant with anomaly mediation
(AMSB). A model of combining MSSM gaugino mediation and anomaly
mediation has been proposed in \cite{Kaplan:2000jz}. By considering
the $Z'$ gaugino as a mediator instead, as well as a different
underlying model, our setup and its phenomenological features are very
different.

One immediate question is whether it is natural for these two
mechanisms to be comparable. As we will discuss in detail in
Sec.~\ref{sec:XD}, such a scenario can be achieved in a large class of
models. Here, we will instead summarize the main features of the
spectrum of soft parameters.  The scale of the soft parameters in both
anomaly and $\zp$-gaugino mediation is set by one dimensionful
parameter for each mechanism. For $\zp$-gaugino mediation this
parameter is the $\zp$ gaugino mass $\mzp$. Up to order one
dimensionless parameters and logarithms of the ratio of the
supersymmetry (SUSY) breaking scale $\Lambda_S$ to $\mzp$, the soft
scalar masses ($m^2_{\tilde{f}}$) and the gaugino masses ($M_a$) are
given by
\begin{eqnarray*}
(m^2_{\tilde{f}})_{{\tilde Z}^\prime{\rm MSB}} \sim  \frac{\mzp^2}{16 \pi^2}, \qquad \qquad
M_a\sim \frac{\mzp}{(16\pi^2)^2}.
\end{eqnarray*}
For AMSB the dimensionful parameter is the gravitino mass $m_{3/2}$.
Very loosely we can write
\begin{eqnarray*}
(m^2_{\tilde{f}})_{\rm AMSB} \sim  \frac{m^2_{3/2}}{(16\pi^2)^2}, \qquad\qquad
M_a\sim \frac{m_{3/2}}{16\pi^2}.
\end{eqnarray*}
Here, we would like to consider a scenario in which contributions to
the soft scalar masses from these two scenarios are comparable. In
this case, the positive contribution from $\zp$ gaugino mediation can
solve the tachyonic slepton mass problem of anomaly mediation. The
gaugino masses, dominated by anomaly mediation, are also of the same
order of magnitude. Therefore, this scenario solves the fine-tuning
problem of $\zp$-gaugino mediation.  We demand
\begin{equation}
\label{eq:ratio}
(m^2_{\tilde{f}})_{{\tilde Z}^\prime{\rm MSB}} \sim (m^2_{\tilde{f}})_{\rm AMSB} \ \Rightarrow \ 
r\equiv \frac{m_{3/2}}{\mzp } \sim 4 \pi,
\end{equation}
i.e., the $\zp$ gaugino mass should be about an order of magnitude
smaller than the gravitino mass. If such a hierarchy holds, the $\zp$
contribution to the MSSM gaugino masses is
$$
M_a\sim \frac{\mzp}{(4\pi)^4}\sim \frac{m_{3/2}}{(4\pi)^5}\ll
\frac{m_{3/2}}{(4\pi)^2},
$$ 
i.e., three order of magnitude suppressed compared to the anomaly
contribution and completely negligible. Again, we will leave the
question of whether such a mild hierarchy between the $\zp$ gaugino
and the gravitino mass can be realized naturally in models to the
discussion in Section \ref{sec:XD}.

As an immediate consequence of having comparable contributions to the
scalar masses from $Z'$-gaugino and anomaly mediation, the tachyonic
slepton problem of pure anomaly mediation is overcome. Another
challenge in anomaly mediation is to obtain the correct ratio of $\mu
/ B_{\mu} $. In the scenario with a $U(1)'$, it is natural to consider
a next-to-minimal supersymmetric standard model (NMSSM) - like
scenario where a tree level $\mu$ term is forbidden by $U(1)'$
symmetry.  This includes most of the supersymmetric $U(1)'$ models
other than those based on $B-L$.  The $U(1)'$ gauge symmetry breaking,
the effective $\mu$ and $B_{\mu}$ parameters, and electroweak symmetry
breaking are all generated dynamically. Although not necessarily a
natural solution for the $\mu/B_{\mu}$ problem, we found that it is
not difficult to find model points which admit successful electroweak
symmetry breaking.

Adding additional $U(1)'$ contributions to anomaly mediation has been
considered before in the literature \cite{Jack:2000cd,
  Murakami:2003pb,Sundrum:2004un}, including the possibility of
combining anomaly and $\zp$ mediation in the context of
$U(1)^\prime_{B-L}$~\cite{Kikuchi:2008gj}. However, our scenario is
different either in the way the $U(1)'$ contribution arises, how the
$\mu/B_{\mu}$ problem is addressed, or in our consideration of the
issue of generating the hierarchy between the gravitino and the $\zp$
gaugino mass from microscopic considerations.

The structure of the rest of the paper is as follows. In section
\ref{sec:XD} we discuss how the required hierarchy between the
gravitino and the $\zp$ gaugino can be obtained in extra-dimensional
models. In section \ref{sec:general} we discuss in general terms a
specific implementation of the joint scenario, combining anomaly
mediation with the model described in the original $\zp$-mediation
papers. In section \ref{sec:numerical} we present the detailed
spectrum for two illustrative points in parameter space.  In section
\ref{sec:conclusions} we present our conclusions. Most of the detailed
expressions are relegated to the appendices.

\section{$Z'$-gaugino mediation and anomaly mediation}\label{sec:XD}

In this section we will show that the mild hierarchy between the $\zp$
gaugino and the gravitino can be obtained if we consider an
extra-dimensional implementation of the model. We begin with a setup
used in the original proposal of gaugino mediation
\cite{Kaplan:1999ac,Chacko:1999mi}.  We assume there is only one flat
extra dimension, $y \in [0,L]$, where the MSSM matter fields are
localized at $y=0$, and the hidden sector responsible for the
supersymmetry breaking is localized on a spatially separated brane at
$y=L$.  Unlike the standard gaugino mediation, we assume that the MSSM
gauge supermultiplets are localized together with the matter fields on
the brane at $y=0$.  The $\zp$ gaugino, on the other hand, propagates
in the bulk. Therefore, a $\zp$ gaugino mass is generated via a direct
coupling to the hidden sector brane, while the $SU(3)_C\times
SU(2)_L\times U(1)_Y$ gauginos remain massless at tree level or their
mass arises from a higher order term. There are several possible
couplings between the $Z'$ and the fields on the hidden sector
brane. We consider the simplest possibility, a brane localized term of
the form
\begin{equation}\label{eq:brane}
 c\,\int d^2\theta\, \frac{X}{M_*^2}\,W_{\tinymath{Z^\prime}}\,W_{\tinymath{Z^\prime}}\delta(y-L),
\end{equation}
where $W_{\tinymath{Z^\prime}}$ is the $\uonep$ field strength, $X$ is
the field whose $F$ component generates the gaugino mass, $L$ is the
size of the extra dimension, $M_*$ is the 5D Planck mass and $c$ is a
constant. The 5D and the 4D Planck masses are related by
$M_*^3\,L=M_P^2$.

When the field $X$ develops an $F$ term, a $\zp$ gaugino mass is
generated,
\begin{equation}
\mzp=c\,\frac{F_X}{M_*^2\,L},
\end{equation}
where the extra factor of $L$ arises from the fact that the wave
function of the zero mode of the $\zp$ gaugino is spread over the
extra dimension.  The gravitino mass is of the order
\begin{equation}
m_{3/2}\sim\frac{F}{M_P}=\frac{F}{\sqrt{M_*^{3}\,L}}.
\end{equation}
If we assume that $F$ and $F_X$ are comparable, we have $M_*L\sim
c^2r^2$, where $r$ is the ratio of the gravitino to the $\zp$ gaugino
mass defined in \refeq{eq:ratio}.  This product of the 5D Planck mass
and the size of the extra dimension is constrained both from above and
below.  Naive dimensional analysis \cite{nda} relates the
compactification scale $L^{-1}$ and the cut-off $\sim M_*$ as
\cite{Kaplan:1999ac,Chacko:1999mi}
\begin{equation}\label{eq:upper}
M_*L\lesssim 16 \pi^2.
\end{equation}
One of the central ingredients of anomaly mediation and gaugino
mediation is to suppress contact terms of the form $1/M_*^2\int
d^4\theta\,Y^\dagger\,Y\,Q_i^\dagger\,Q_j,$ with $Y (Q)$ hidden
(visible) sector fields, which can potentially violate flavor
constraints. This is the so-called sequestering. It has been argued
\cite{Randall:1998uk,Luty:1999cz} that locality in the extra dimension
can gives rise to exponential suppression $\sim e^{- m_Y L}$. Taking
$m_Y \sim M_*$, the constraints on first two generation flavor
changing neutral currents lead to \cite{Kaplan:2000jz}
\begin{equation}\label{eq:lower}
M_*L\gtrsim 16.
\end{equation}
The conditions in \refeq{eq:upper} and \refeq{eq:lower} imply that 
\begin{equation}\label{eq:allowed}
4\lesssim c\, r\lesssim 4 \pi.
\end{equation}
As we have discussed in the previous section, we need $r \sim
\mathcal{O}( 10)$ for $\zp$-gaugino mediation and anomaly mediation to
give comparable contributions to the soft scalar masses.  With an
order one coefficient in \refeq{eq:brane} we can easily generate the
appropriate mass hierarchy.

We remark that the actual extra-dimensional model is likely to have
additional structure. In particular, it has been argued
\cite{Luty:2000ec,Anisimov:2001zz,Kachru:2006em,Kachru:2007xp} that
warped compactification could be necessary for successful
sequestering.

We will not go into details of building a realization of our scenario
in a warped space. We only comment that most of the relevant features
of the flat extra-dimensional model do not change significantly since
they are mainly determined by the physics below the compactification
scale. The phenomenological study presented later in this paper begins
with a general parameterization of the boundary condition of
supersymmetry breaking, and is not specific to any particular kind of
compactification.

We would like to make a more detailed comparison with the scenario
studied in Ref.~\cite{Kaplan:2000jz}. In addition to anomaly
mediation, MSSM gauginos are employed as the mediators of
supersymmetry breaking. In that case, operators of the form of
\refeq{eq:brane}, with obvious substitution of the $Z'$ field strength
superfield with the corresponding ones for the MSSM gauge fields, give
the dominant contributions to the MSSM gauginos in comparison with the
anomaly mediation contribution, unless $M_* L \sim 10^4$ which is
difficult to realize. To avoid that, such a coupling has to be absent
and additional higher-order interactions lead to comparable
contributions from gaugino mediation and anomaly mediation. In our
case, while the operator of \refeq{eq:brane} gives the dominant
contribution to the $Z'$ gaugino mass, our key requirement is that the
contributions to the scalar mass are comparable. As we have already
seen, this is a much milder condition and easy to satisfy.  The MSSM
gaugino masses are then almost completely determined by anomaly
mediation.

We also emphasize that the scenario we have presented in this section
is a specific implementation of the more general $\zp$-gaugino
mediation scenario of \cite{Langacker:2007ac,Langacker:2008ip}, and
the two are not equivalent. In general, we can consider a different
scenario of $Z'$ gauge mediation, in which the $Z'$ couples to a
messenger sector and the boundary condition is different from
\refeq{eq:brane}. The $Z'$ gaugino and scalar charged under $U(1)'$
receive supersymmetry breaking masses at the same order
\cite{Langacker:1999hs}. This is an interesting possibility which we
will not pursue further.  We also note that in the extra-dimensional
setup, one can consider a scenario in which the operator in
\refeq{eq:brane} is absent, and the $\zp$ couples to a messenger
sector on the hidden sector brane.  It was pointed out in
\cite{Kaplan:1999ac} that, even in this case, the boundary values of
the scalar masses are still small in comparison with the gaugino mass,
and the low energy spectrum is still that of the gaugino mediation.

\section{Specific implementation: General Expressions }\label{sec:general}
Having shown that combining $\zp$ and anomaly mediation is natural
within this class of extra-dimensional models, we now present an
explicit implementation. We choose to do that using the same model as
in the original $\zp$-mediation papers
\cite{Langacker:2007ac,Langacker:2008ip}. While certainly not the only
possible realization, it is probably one of the simplest
possibilities.
\subsection{The model}
\begin{itemize}
\item We introduce a new $\uonep$ gauge symmetry under which all the
  MSSM fields are charged. The $\uonep$ charges are family universal.
\item The charges of $H_u$ and $H_d$ are such that an elementary $\mu$
  term in the superpotential is not allowed. Instead we introduce a SM
  singlet superfield $S$ which is charged under $\uonep$, such that
  the superpotential term $S H_u H_d$ is allowed.
\item To cancel the new anomalies we introduce the following ``exotic"
  matter:
\begin{itemize}
\item 3 pairs of colored, $SU(2)_L$ singlet exotics $D,D^c$ with
  hypercharge $Y_D=-1/3$ and $Y_{D^c}=1/3$.
\item 2 pairs of uncolored $SU(2)_L$ singlet exotics $E,E^c$ with
  hypercharge $Y_E=-1$ and $Y_{E^c}=1$.
\end{itemize}
\item The exotic fields can couple to $S$, namely the superpotential
  terms $S D D^c$ and $S E E^c$ are allowed. 
\item Normalizing the charges by $Q_{H_d}=1$, $Q_{H_u}$ and $Q_Q$ are
  the only free parameters.  The other charges are determined by the
  anomaly cancellation conditions and the allowed superpotential
  couplings. The explicit relations are listed in appendix
  \ref{app:charges}.
\end{itemize}

The superpotential is
\begin{eqnarray}
W&=& y_u  {H}_u {Q} {u}^c + y_d  {H}_d {Q} {d}^c + y_e  {H}_d {L} {e}^c+y_\nu  {H}_u {L} {\nu}^c
\\\nonumber
&+&\lambda  {S}  {H}_u  {H}_d+ y_D\, S\left( \sum_{i=1}^{3}  D_i D_i^c \right) +y_E\, S\left (\sum_{j=1}^{2} E_j E_j^c\right).
\end{eqnarray}

We assume that the $\zp$ gaugino mass is generated at the SUSY
breaking scale $\Lambda_S$. The other gauginos and scalar masses at
$\Lambda_S$ are generated from the anomaly contribution.  We use the
general expressions from \cite{Gherghetta:1999sw}, collected in
Appendix~\ref{app:bc} for completeness.

One interesting feature of this model is that the $\beta$-function of
the strong coupling vanishes at one-loop order.  The gluino mass is
generated almost exclusively by the anomaly contribution, which is
proportional to this $\beta$-function. This implies that the gluino
mass is zero at one loop, but will get a non-zero contribution at the
two-loop level. Nevertheless, its size can be comparable to the wino
and bino masses, which are non-zero already at one-loop order. In
particular the two-loop gluino mass is still much larger than the
$\zp$ contribution.  As a result one finds that for a generic choice
of parameters the gaugino mass hierarchy is $M_1>M_3>M_2$. This should
be compared to the ``standard" AMSB for which the gaugino mass
hierarchy is such that the gluino is heavier than the bino and the
wino. In other words, since we are considering a non-minimal extension
of the standard model, the hierarchy of the gauge coupling
$\beta$-functions, and consequently the gaugino mass hierarchy, is
different from that of the MSSM. For consistency we will calculate all
of the MSSM gaugino masses at two-loop level.

The effect of the $\zp$ gauge coupling $\beta$-function must also be
included in the anomaly contribution to the scalar masses. Therefore,
compared to the standard AMSB, it is possible to find that more
scalars apart from the sleptons are tachyonic at the UV boundary.
Vacuum stability in the very early universe could constrain such a
scenario~\cite{Linde:1981zj}. However, without a compelling model of
that era of cosmic evolution, we will not take this as a constraint on
our parameter space.

We would like to emphasize that the vanishing of the strong coupling
$\beta$-function at one-loop order is not an accident, but a rather
general result following from these assumptions: introduce $n_D$
generations of exotic quarks (i.e., $n_D$ triplets and $n_D$
anti-triplets), demanding that there is a single SM singlet field $S$
(or a set of $S$ fields with the same $\uonep\,$ charge)\footnote{This
is {\em not} the case for the minimal gauge unification models
considered in \cite{Erler:2000wu,Langacker:2008dq}.}  which generates
the effective $\mu$ term and gives mass to the exotic quarks, and
allow the standard quark Yukawa couplings. The cancellation of the
$SU(3)^2_C\times\uonep$ anomaly then requires that $n_D=3$
\cite{Langacker:2008ip}. As a result the strong coupling $\beta$
function vanishes at one loop.

While the boundary condition at the SUSY breaking scale for
$M_\lambda, m_{\tilde{f}_i}$ and $A_y$ arise only from the anomaly
contribution, the renormalization group equations (RGEs) for these
parameters also receive contributions from the interactions with the
$\zp$ gaugino.  We run the RGEs down to the electroweak (EW) scale
using the two-loop RGEs for the gauge couplings and the gaugino
masses, and the one-loop RGEs for all the other parameters.  The
explicit formulas for the RGEs are listed in appendix
\ref{app:RGEs}. The RGEs can be solved numerically.

Around the EW scale we minimize the scalar potential for the neutral
Higgses and the scalar component of $S$. It is given by
\cite{Langacker:2008yv}
\begin{eqnarray}
V(S,H_u^0,H_d^0)    &=&m_S^2|S|^2+m^2_{H_u}|H_u^0|^2+m^2_{H_d}|H_d^0|^2+\nonumber\\
        &+&|\lambda|^2\left(|S|^2|H_u^0|^2+|S|^2|H_d^0|^2+|H_u^0|^2|H_d^0|^2\right)-(ASH_u^0H_d^0+\,{\rm h.c})\nonumber\\
        &+&\frac18\,(g_2^2+\frac35 g_1^{2})\left(|H_u^0|^2-|H_d^0|^2\right)^2+
        \frac12\gz^2\left(Q_{H_u}|H_u^0|^2+Q_{H_d}|H_d^0|^2+Q_S|S|^2\right)^2,\nonumber\\
\end{eqnarray}
where $g_2$ is the $SU(2)_L$ gauge coupling and $g_1$ is related to
the hypercharge gauge coupling $g_Y$ via $g_1^2=5g^2_Y/3$. The vacuum
at this scale should break the $U(1)'$ symmetry as well as the
electroweak symmetry. We typically require that the vacuum expectation
value (vev) of $S$ is larger than the EW scale.  Using these vevs we
can calculate the spectrum.

\subsection{The spectrum calculation}
In this section, we review the method we employed to calculate the low
energy spectrum from the UV inputs.  Several of the mass matrices
needed for the calculation of the spectrum can be found in the
literature:
\begin{itemize}
\item Those for the $Z$ and $Z^\prime$ gauge bosons and the
  neutralinos can be found in \cite{Langacker:2008yv}.
\item Those for the charginos are the same as in the MSSM, see for
  example \cite{Martin:1997ns}, but with $\mu$ replaced by $\lambda
  \langle S \rangle$.
\item The tree level expressions for the scalar mass matrices of $H_u,
  H_d$ and $S$ can be found in \cite{Barger:2006dh}.
\end{itemize}
The masses of the fermions are given by
\begin{equation}
m_{f_i}=y_i\langle \phi_i \rangle.
\end{equation}
where $y_i$ are non-negligible only for the top quark, the $b $ quark,
the $\tau$ lepton, and the exotics $D$ and $E$. Also, $\phi_t=H_u^0$,
$\phi_{b,\tau}=H_d^0$, and $\phi_{D,E}=S$.  All the vevs are assumed
to be real.

The sfermion mass matrices can be written in a compact form as
\begin{equation}\label{eq:mass}
m_{\tilde{f}_i}^2=\left(\begin{array}{ccc}
m^2_{\tilde{f}_{i1}}+y_i^2 \langle \phi_i \rangle^2+\Delta_{{\tilde
f}_{i1}}&\hspace{3em}&\pm \lambda\,y_i\langle S \rangle \langle
H_u^0\rangle \langle H_d^0\rangle/\langle \phi_i \rangle
\mp A^*_{y_i}\langle \phi_i \rangle\\
 \pm \lambda\,y_i\langle S \rangle
\langle H_u^0\rangle\langle H_d^0\rangle/\langle \phi_i
\rangle \mp A_{y_i}\langle \phi_i
\rangle&\hspace{3em}&m^2_{\tilde{f}_{i2}}+y^2_i \langle \phi_i
\rangle^2+\Delta_{{\tilde f}_{i2}}
\end{array}
\right),
\end{equation}
where the upper signs are for the $\tilde{t}$ and the lower are for
the $\tilde{b}$, $\tilde{\tau}$, $\tilde{D}$, and $\tilde{E}$.  In the
last equation $y_i$ and $A_{y_i}$ are non-zero only for the stops,
sbottoms, staus, and the scalar exotics. The notation for the soft
masses is such that for squarks and sleptons $m^2_{\tilde{f}_{i1}}$ is
the $SU(2)_L$ doublet soft mass and $m^2_{\tilde{f}_{i2}}$ is the
right-handed soft mass.  For the exotics
$m^2_{\tilde{f}_{D1}}=m^2_{\tilde
D},\,m^2_{\tilde{f}_{E1}}=m^2_{\tilde E},\,
m^2_{\tilde{f}_{D2}}=m^2_{\tilde{D}^c},\,m^2_{\tilde{f}_{E2}}=m^2_{\tilde{E}^c}$.
Also
\begin{eqnarray}
\Delta_{{\tilde
f}_{i1}}&=&\frac12\left(g^2_2T^3_{i1}-\frac35g_1^2Y_{i1}
\right)\left(\langle H_d^0\rangle^2-\langle H_u^0\rangle^2\right) +
\gz^2 Q_{i1}\left(Q_S\langle S \rangle^2+Q_{H_u}\langle H_u^0\rangle^2+Q_{H_d}\langle H_d^0\rangle^2 \right)\nonumber\\
\Delta_{{\tilde
f}_{i2}}&=&\frac12\left(g^2_2T^3_{i2}-\frac35g_1^2Y_{i2}
\right)\left(\langle H_d^0\rangle^2-\langle H_u^0\rangle^2\right) +
\gz^2 Q_{i2}\left(Q_S\langle S \rangle^2+Q_{H_u}\langle
H_u^0\rangle^2+Q_{H_d}\langle H_d^0\rangle^2 \right),\nonumber\\
\end{eqnarray}
where $T^3$ and $Y$ are the third component
of the weak isospin and the weak hypercharge, respectively. 

There are also important one-loop radiative corrections to the scalar
masses of $H_u, H_d$ and $S$.  These are most easily calculated by
using the one-loop Coleman-Weinberg potential \cite{Coleman:1973jx}.
We limit ourselves to the effects of the stop and top loops, which due
to the large top Yukawa coupling are expected to be the dominant
ones. The one-loop effective potential in the $\overline{{\rm
DR}^\prime}$ scheme is \cite{Barger:2006dh,Martin:2001vx}
\begin{equation}
V^1=\frac3{32\pi^2}\left[\sum^2_{j=1}m^4_{\tilde{t}_j}\left(\ln\frac{m^2_{\tilde{t}_j}}{\mu^2}-\frac32\right)-
2\bar{m}^4_t\left(\ln\frac{\bar{m}^2_t}{\mu^2}-\frac32\right)\right],
\end{equation}
where $\mu$ is the renormalization scale, $\bar{m}^2_t=y^2_t|H^0_u|^2$,
and $m^2_{\tilde{t}_j}$ are the field-valued eigenvalues of
(\ref{eq:mass}) for the case of stops.

Let $\langle\phi_i\rangle$ be the vev of $H_u^0$,$H_d^0$, or $S$
\emph{including} the effect of the radiative corrections. We now
expand each field around its vev as
\begin{equation}
\phi_j=\langle\phi_j\rangle+\frac1{\sqrt{2}}\phi^R_j+i\frac1{\sqrt{2}}\phi^I_j.
\end{equation}

The one-loop corrections for the mass matrices of the CP even (${\cal
  M}^1_+$) and CP odd (${\cal M}^1_-$) ``Higgses" can be written as
\cite{Barger:2006dh}
\begin{eqnarray}\label{eq:massrad}
\left({\cal M}^1_{+}\right)_{ij}&=&\frac{\partial^2 V^1}{\partial\phi^R_i\,\partial\phi^R_j}\bigg|_0-
\delta_{ij}\frac1{\sqrt{2}\langle\phi_i\rangle}\frac{\partial V^1}{\partial\phi^R_i}\bigg|_0\nonumber\\
\left({\cal M}^1_{-}\right)_{ij}&=&\frac{\partial^2 V^1}{\partial\phi^I_i\,\partial\phi^I_j}\bigg|_0-
\delta_{ij}\frac1{\sqrt{2}\langle\phi_i\rangle}\frac{\partial V^1}{\partial\phi^R_i}\bigg|_0.
\end{eqnarray}
Instead of displaying explicit analytical results for these mass
matrices, it is easier to calculate them for a given point in
parameter space. In calculating the radiative corrections we do not
set the various gauge coupling to zero, since the $D$-term
contributions to the stop masses can be quite substantial. From
(\ref{eq:massrad}) we obtain the masses for the ``Higgs" particles
listed in section \ref{sec:numerical}.  Finally, the radiative
corrections to the charged Higgs masses can be determined by $SU(2)_L$
symmetry.

\section{Specific implementation: Illustration points}\label{sec:numerical}
To show that the model can lead to a reasonable spectrum we choose two
specific illustration points. We list the input parameters and the
resulting spectrum. We have checked that all the masses for the
supersymmetric particles are allowed by the current experimental
bounds.

The input parameters can be divided into dimensionless and
dimensionful parameters. The dimensionless input parameters are the
$U(1)'$ charges of $H_u$ and $Q$ (the quark doublet), which are listed
in Table~\ref{tab:charges}, the $U(1)'$ gauge coupling $\gz$, and the
superpotential couplings $y_t,y_b,y_\tau,y_D,y_E$ and $\lambda$. The
dimensionless couplings are chosen to be
\begin{eqnarray}
\uonep \mbox{ gauge coupling (at } \Lambda_{S})&:&\quad \gz=0.45\nonumber\\
\mbox{Superpotential parameters (at } \Lambda_{\rm EW})
&:&\quad\lambda=0.1,\, y_D=0.3,\, y_E=0.5.
\end{eqnarray}
The values of $y_t,\,y_b$ and $y_\tau$ will be chosen below such that
at the electroweak scale they reproduce the values of the top quark
mass \cite{Amsler:2008zzb}, the $b$ quark mass \cite{Abdallah:2005cv},
and the $\tau$ lepton mass \cite{Amsler:2008zzb}, where we ignore the
small running effect on the $\tau$ lepton mass.

\begin{table}[!ht]
\begin{center}
\begin{tabular}{c | c c c c c c c c c c c c c }
\toprule
\multicolumn{1}{c}{}&$Q$&$u^c $&$d^c$&$L$&$\nu^c$&$e^c$&$H_u$&$H_d$&$S$&$D_i$&$D_i^c$&$E_i$&$E_i^c $\\
\midrule
$Q_i$&$-\frac 13 $&$\frac {11}{15}  $&$-\frac 23 $&
$\frac 45 $&$-\frac 2 5 $&$-\frac 9 5 $&$-\frac 25 $
&$1 $&$-\frac 3 5 $&$\frac 45 $&$-\frac 15 $&$\frac 9 5 $&$-\frac 65 $\\
\bottomrule
\end{tabular}
\caption{$U(1)^\prime$ charges used in the model. $Q_{H_d}$ is
  normalized to 1. We have chosen $Q_{H_u}=-\frac25$, and
  $Q_Q=-\frac13$. The rest of the charges are determined by anomaly
  cancellation and gauge invariance.}
\label{tab:charges}
\end{center}
\end{table}

\begin{figure}[!ht]
\begin{center}
\includegraphics[width=8.25cm]{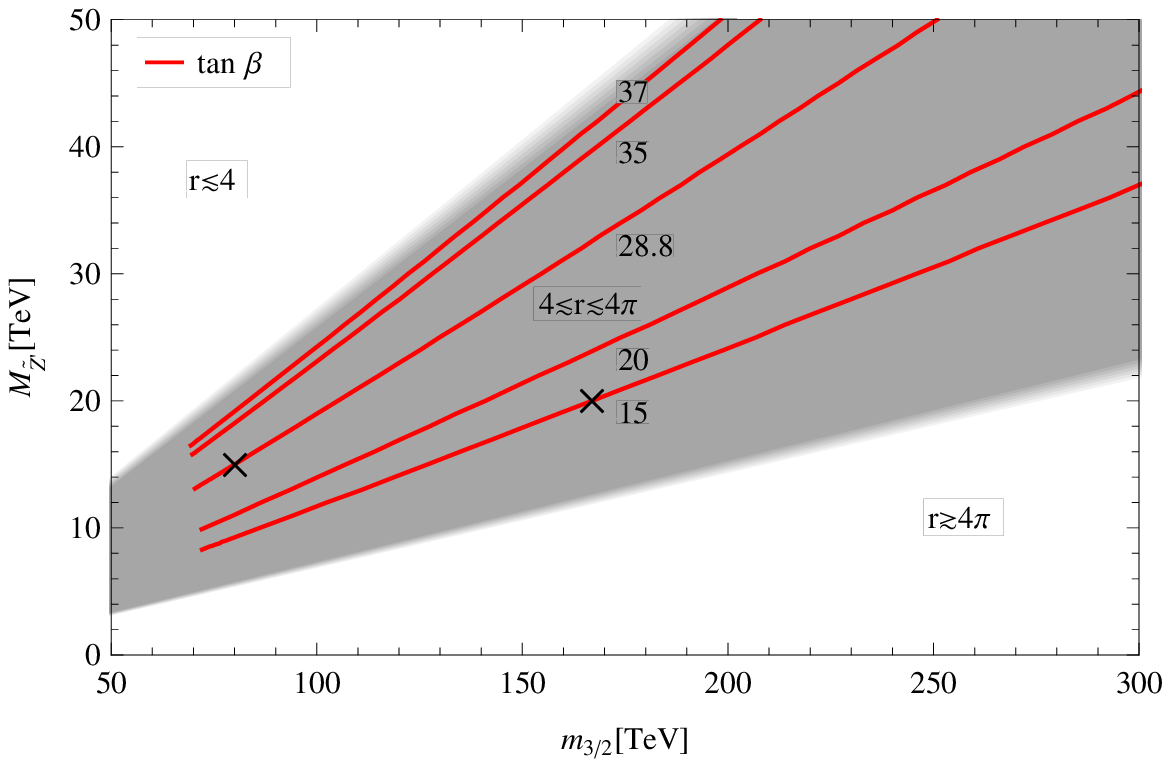}
\includegraphics[width=8.25cm]{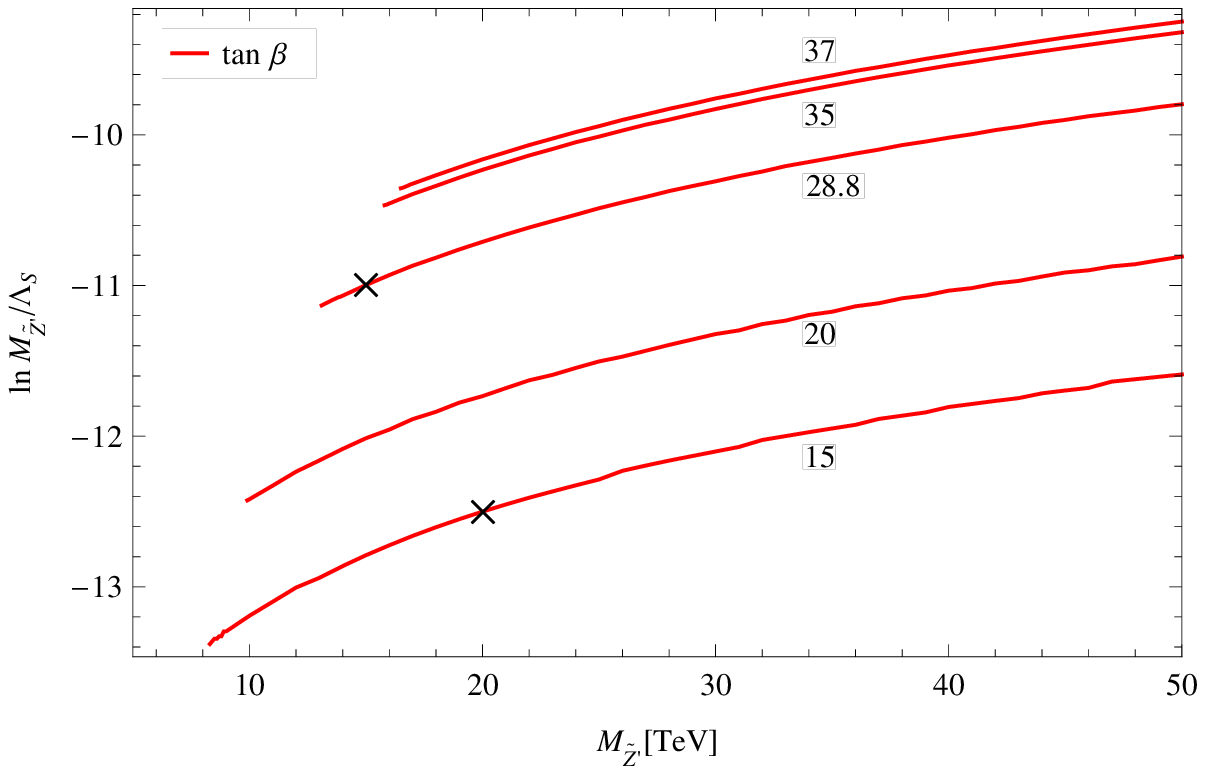}
\caption{Constant $\tan{\beta}$ lines in the $\mzp-m_{3/2}$ plane
  (left) and the corresponding values of $\Lambda_S$ along such lines
  as a function of $\mzp$(right). The shaded region in the left plot
  corresponds to the allowed range in \refeq{eq:allowed}
  ($r=m_{3/2}/M_{{\tilde Z}^\prime}$) for $c=1$. The position of our
  two illustration points is indicated by the crosses.
\label{constanb}}
\label{fig:constanb}
\end{center}
\end{figure}

The dimensionful input parameters are the gravitino mass $m_{3/2}$,
the $\zp$ gaugino mass $\mzp$ and the SUSY breaking scale
$\Lambda_S$. They must be chosen such that the electroweak and
$U(1)^\prime$ symmetry breaking occurs when we run down to the EW
scale. We also demand that the ratio of the gravitino mass to the
$\zp$ gaugino is in the allowed range of \refeq{eq:allowed}, with
$c=1$.  The values of $\tan \beta$ compatible with a realistic
spectrum for our choice of the charges, $\gz$, $\lambda$, $y_D$ and
$y_E$ can be read from Fig.~\ref{constanb} where the lines of constant
$\tan{\beta}$ are drawn as a function of the scales of the model. We
find
$$15\lesssim\tan{\beta}\lesssim 37.$$
Moving along each line towards higher values of the $\zp$-gaugino and
gravitino masses the overall spectrum is heavier.

%
%

\begin{figure}[!ht]
\begin{center}
\includegraphics[width=10cm]{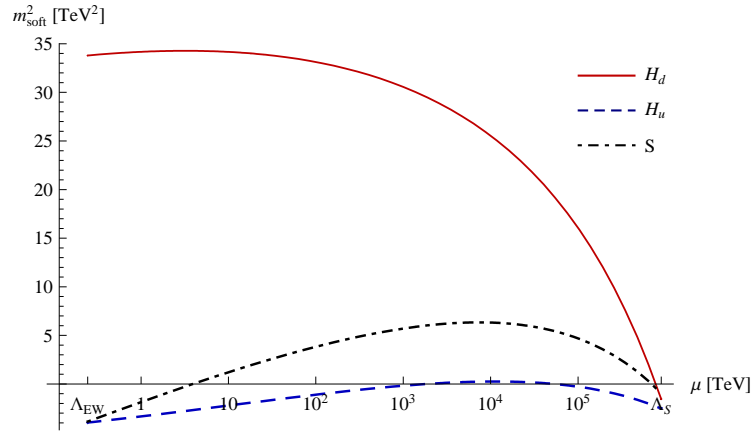}
\caption{Scalar soft masses squared for the Higgs fields in the first illustration point.} 
\label{fig:higgses}
\end{center}
\end{figure}

We choose two illustration points. The first has
\begin{equation}
m_{3/2}=80\, {\rm TeV}, \quad \mzp=15\,  {\rm TeV}, \quad \Lambda_S\sim  10^9\, {\rm GeV}.
\end{equation}
The top, bottom, and tau Yukawa couplings are taken to be
\begin{equation}
y_t = 1, \, y_b = 0.5,\,y_\tau = 0.294
\end{equation}
at the EW scale. For the first illustration point the vacuum
parameters are
\begin{equation}
\tan\beta=28.8,\quad \langle S\rangle=11.9 \,{\rm TeV}.
\end{equation}
In Fig.~\ref{fig:higgses} we show the running of the scalar
soft masses of $H_u$, $H_d$ and $S$, for this point.

%

The second illustration point has 
\begin{equation}
m_{3/2}\approx 167\, {\rm TeV}, \quad \mzp=20\,  {\rm TeV}, \quad \Lambda_S\sim  5\cdot 10^9\, {\rm GeV}.
\label{eq:scales2}
\end{equation}
The top, bottom, and tau Yukawa couplings are taken to be
\begin{equation}
y_t = 1, \, y_b = 0.257,\,y_\tau = 0.154
\end{equation}
at the EW scale. The vacuum
parameters are
\begin{equation}
\tan\beta=15,\quad \langle S\rangle=25.2 \,{\rm TeV}.
\end{equation}

In the Tables \ref{tab:higgs} to \ref{tab:Zp_boson} we display the
details of the spectrum for both points. We use white and gray
background colors to distinguish between the first and second points,
respectively.

Due to the large $\tan\beta$ and the large vev of $S$, the vevs are
strongly ordered : $\langle H_d^0\rangle \ll\langle
H_u^0\rangle\ll\langle S\rangle$.  As a result there is very little
mixing in the extended Higgs sector ($H_u$,$H_d$,$S$).  The Higgs
masses, including the one-loop radiative corrections, and their
composition are listed in Table~\ref{tab:higgs}.
%
\begin{table}[!ht]
\begin{center}
{\small
\begin{tabular}{c c c | c c c c c c }
\ctoprule
&\multicolumn{2}{c}{\underline{$m$ [TeV]}}\vline&
\multicolumn{6}{c}{\underline{Composition[$\%$]}}\\
$ $&$ $&$
$&\multicolumn{2}{c}{$H_d$}&\multicolumn{2}{c}{$H_u$}&\multicolumn{2}{c}{$S$}\\
\midrule
$h^0 $&0.138&\graycell 0.142&0.1&\graycell 0.4 & 99.9 &\graycell 99.6 & 0 &\graycell 0 \\
$H^0_1 $& 2.79 &\graycell 5.69 & 0 &\graycell 0 & 0 &\graycell 0 &100 &\graycell 100\\
$A^0 $& 4.78&\graycell 6.85 & 99.9 &\graycell 99.6 & 0.1&\graycell 0.4 & 0 &\graycell 0 \\
$H^0_2 $& 4.78 &\graycell 6.85 & 99.9 &\graycell 99.6 & 0.1 &\graycell 0.4 & 0 &\graycell 0 \\
$H^\pm $& 4.78 &\graycell 6.85 & 99.9 &\graycell 99.6 & 0.1 &\graycell 0.4 & - &\graycell - \\
\cbottomrule
\end{tabular}
}
\caption{Neutral and charged Higgs masses (including radiative
  corrections) and composition, for point 1 (white background) and
  point 2 (gray background).}
\label{tab:higgs}
\end{center}
\end{table}

The radiative corrections can be quite substantial. For example, at
tree level we have $m_{h^0}=85$ GeV, $m_{H^0_1}=2.78$ TeV, and
$m_{H^0_2}=5.37$ TeV for the first illustration point. For the
lightest Higgs boson, two-loop effects typically reduce the one-loop
mass by a few GeV \cite{Hahn:2009zz}. The radiative corrections to the
charged Higgs masses were determined by $SU(2)_L$ symmetry and not by
a direct calculation.

The gluino mass is shown in Table \ref{tab:gaugino_masses}.  We also
include in the table the bino and wino mass parameters to compare with
the standard anomaly mediation scenario. As described in the previous
section, we find that like the standard AMSB the wino is the lightest
of the gauginos, but unlike standard AMSB the gluino is lighter than
the bino.
\begin{table}[!ht]
\begin{center}
{\small
\begin{tabular}{c c | c c | c c }
\ctoprule
\multicolumn{2}{c}{$M_{1}$ [TeV]}\vline&\multicolumn{2}{c}{$M_{2}$ [TeV]}\vline&\multicolumn{2}{c}{$M_{3}$ [TeV]}\\
\midrule
 1.17 &\graycell 2.41 & 0.279 &\graycell 0.582 & 0.399 &\graycell 0.813 \\
\cbottomrule
\end{tabular}
}
\caption{Gaugino mass parameters for point 1 (white background) and
  point 2 (gray background). 
The neutralino and chargino mass eigenvalues are given in Tables \ref{tab:neutralinos} and \ref{tab:charginos}, respectively.}
\label{tab:gaugino_masses}
\end{center}
\end{table}

\begin{figure}[!ht]
\begin{center}
\includegraphics[width=8.25cm]{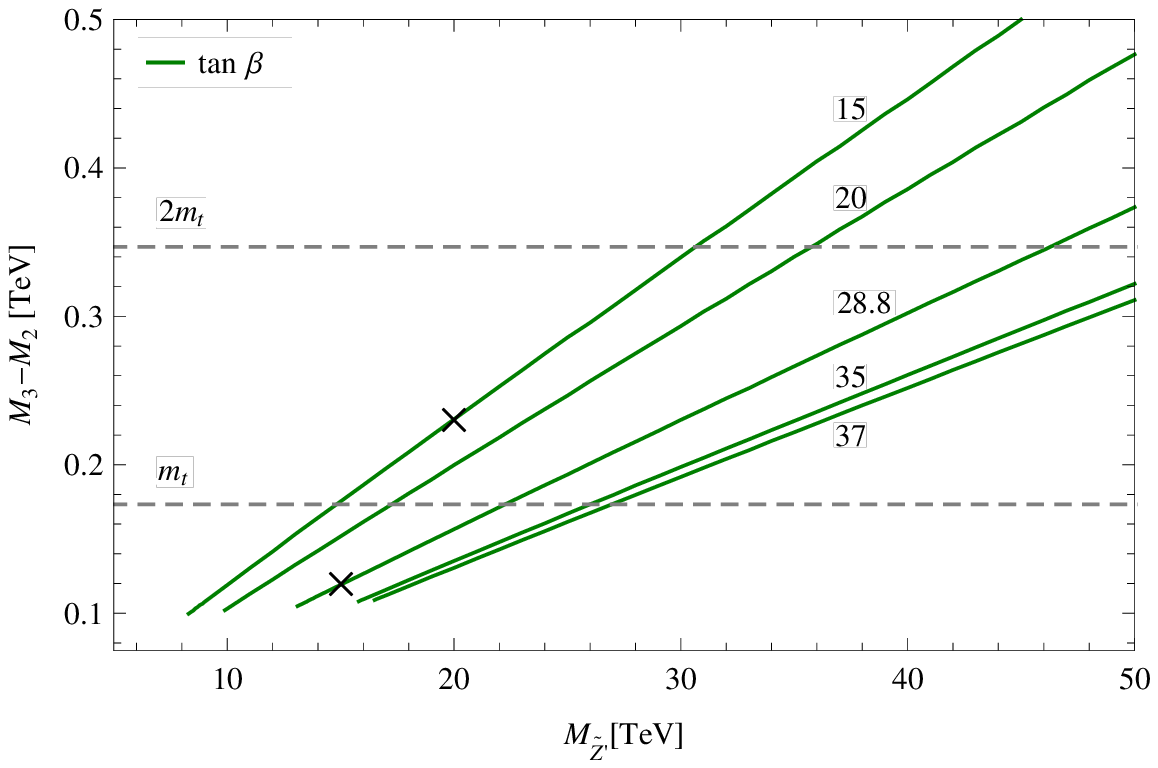}
\includegraphics[width=8.25cm]{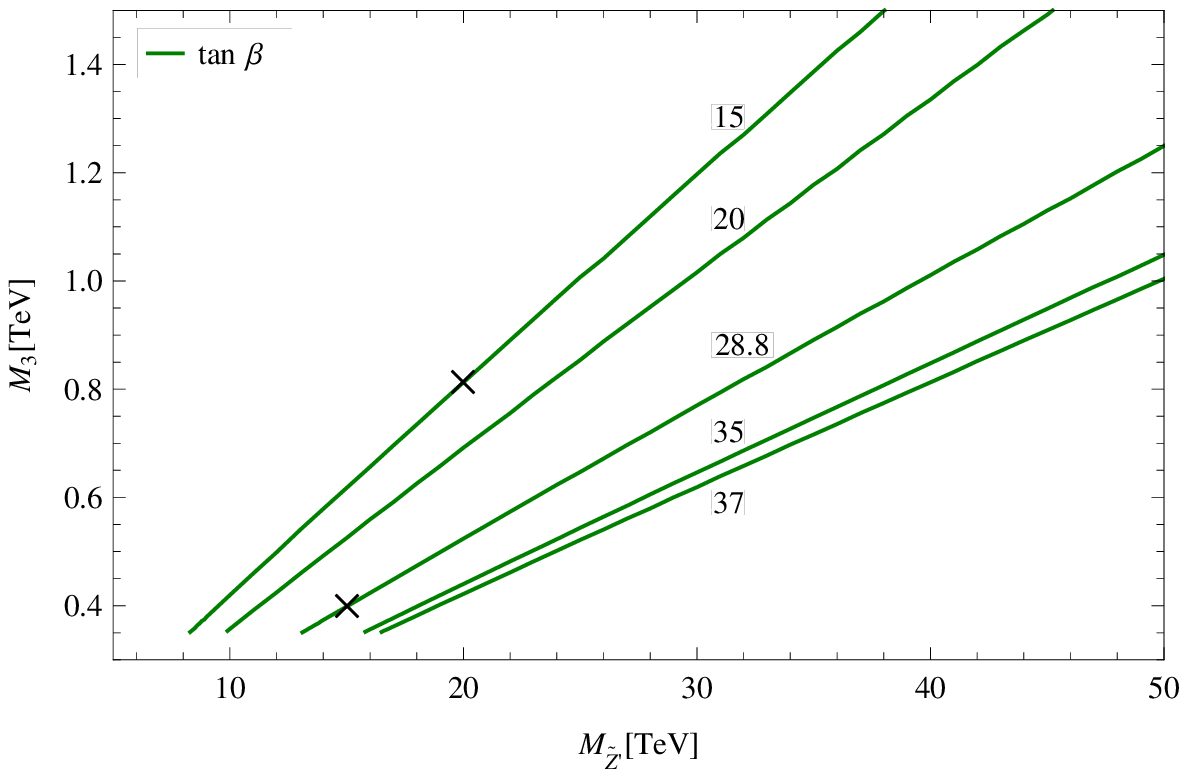}
\caption{(Left) Mass splitting between the gluino and wino masses for
  each $\tan{\beta}$ line in Fig.~\ref{constanb}. (Right) Gluino mass
  along the same lines. Again, the two points discussed in the text
  are indicated by a cross.
\label{M3M2plot}}
\label{fig:M3M2plot}
\end{center}
\end{figure}

As can be seen from Fig.~\ref{constanb}, one can raise the overall
scale and still find acceptable spectrum. One can vary the mass
splitting between $M_3$ and $M_2$, which is approximately the mass
difference between the gluino and the lightest supersymmetric particle
(LSP), by increasing the overall scale. In Fig. \ref{M3M2plot} we show
how such splitting as well as the gluino mass grow as we lift the
$\zp$-gaugino and gravitino masses along each constant $\tan{\beta}$
line in Fig.~\ref{constanb}.  Whether this mass splitting is less than
$m_t$, bigger than $m_t$, or bigger than $2 m_t$, has important
implications to phenomenology, as discussed in detail in section
\ref{sec:conclusions}. The largest splitting for our illustration
points is for the second one, with $M_3-M_2 \sim 230 \ {\rm GeV} $,
and we observe that for gluinos below a TeV we can achieve values up
to the order of 300 GeV. The gluino masses we consider are not
excluded by the jets $+ \not{\!\!E}_T$ search at the Tevatron
\cite{gluino}. It is also possible to probe this scenario at the
Tevatron if the gluino mass is in the range of 300 - 400 GeV.

%
%

Similarly to the Higgs sector, there is generally very little mixing
in the neutralino sector. The only exception are the Higgsinos, which
mix within themselves almost maximally, due to the large effective
$\mu$ term. The neutralino masses and composition are listed in Table
\ref{tab:neutralinos}. In our case, as can be seen especially in the
first point, there is also some Higgsino-bino mixing because of the
small difference between the effective $\mu$ ($=\lambda\langle
S\rangle$) and $M_1$.
\begin{table}[!ht]
\begin{center}
{\small
\begin{tabular}{c c c | c c c c c c c c c c c c}
\ctoprule
&\multicolumn{2}{c}{\underline{$m_{\tilde{N}}$ [TeV]}}\vline& \multicolumn{12}{c}{\underline{Composition[$\%$]}}\\
$ $&$ $&$ $&\multicolumn{2}{c}{${\tilde B}$}&\multicolumn{2}{c}{${\tilde W}^0$}&\multicolumn{2}{c}{${\tilde \zp}$}&\multicolumn{2}{c}{${\tilde H}_d^0$}&\multicolumn{2}{c}{${\tilde H}_u^0 $}&\multicolumn{2}{c}{${\tilde S} $} \\
\midrule
${\tilde N}_1 $& 0.278 &\graycell 0.582 & 0 &\graycell 0 & 99.5 &\graycell 99.9 & 0 &\graycell 0 & 0.5 &\graycell 0.1& 0 &\graycell 0 & 0 &\graycell 0 \\
${\tilde N}_2 $& 0.612 &\graycell 1.81 & 0 &\graycell 0 & 0 &\graycell 0 & 4.6 &\graycell 9.2 & 0 &\graycell 0 & 0.1 &\graycell 0 & 95.3 &\graycell 90.7 \\
${\tilde N}_3 $& 1.15 &\graycell 2.41 & 71.2 &\graycell 94.1 & 0.1 &\graycell 0 & 0 &\graycell 0 & 14.9 &\graycell 3.1 & 13.7 &\graycell 2.8 & 0 &\graycell 0 \\
${\tilde N}_4 $& 1.19 &\graycell 2.52 & 0 &\graycell 0 & 0.2 &\graycell 0 & 0 &\graycell 0 & 49.8 &\graycell 50 & 50 &\graycell 50 & 0 &\graycell 0 \\
${\tilde N}_5 $& 1.21 &\graycell 2.53 & 28.8 &\graycell 5.9 & 0.2 &\graycell 0.1 & 0 &\graycell 0 & 34.8 &\graycell 46.8 & 36.2 &\graycell 47.2 & 0 &\graycell 0 \\
${\tilde N}_6 $& 12.7 &\graycell 17.8 & 0 &\graycell 0 & 0 &\graycell 0 & 95.4 &\graycell 90.8 & 0 &\graycell 0 & 0 &\graycell 0 & 4.6 &\graycell 9.2 \\
\cbottomrule
\end{tabular}
}
\caption{Neutralino masses and composition for point 1 (white background) and
  point 2 (gray background). }
\label{tab:neutralinos}
\end{center}
\end{table}

\begin{table}[!ht]
\begin{center}
{\small
\begin{tabular}{c c c | c c c c }
\ctoprule
&\multicolumn{2}{c}{\underline{$m_{\tilde{C}}$ [TeV]}}\vline& \multicolumn{4}{c}{\underline{Composition[$\%$]}}\\
$ $&$ $&$ $&\multicolumn{2}{c}{${\tilde W}^\pm $}&\multicolumn{2}{c}{${\tilde H}^\pm$}\\
\midrule
${\tilde C}_1 $& 0.278 &\graycell 0.581 & ~~99~~ &\graycell 99.8 & 1 &\graycell 0.2 \\
${\tilde C}_2 $& 1.2 &\graycell 2.52 & 1 &\graycell 0.2 & ~~99~~ &\graycell 99.8 \\

\cbottomrule
\end{tabular}
}
\caption{Chargino masses and composition for point 1 (white background) and
  point 2 (gray background). }
\label{tab:charginos}
\end{center}
\end{table}

Due to the large difference between $M_2$ and the effective $\mu$ ,
the mixing in the chargino sector is also very small. The chargino
spectrum is displayed in Table~\ref{tab:charginos}.  The ${\tilde C}_1
$ is expected to be heavier than the ${\tilde N}_1 $ by $\sim 160$ MeV
due to radiative corrections~\cite{Pierce:1996zz}.

\begin{table}[!ht]
\begin{center}
{\small
\begin{tabular}{c c c | c c c | c c c | c c c }
\ctoprule
\multicolumn{6}{c}{$1^\mathrm{st}$ and $2^\mathrm{nd}$ families}\vline& \multicolumn{6}{c}{$3^\mathrm{rd}$ family}\\
\midrule
$ $&\multicolumn{2}{c}{\underline{$m_{\tilde f}$ [TeV]}}&$ $&\multicolumn{2}{c}{\underline{$m_{\tilde f}$ [TeV]}}\vline&$ $&\multicolumn{2}{c}{\underline{$m_{\tilde f}$ [TeV]}}&$ $&\multicolumn{2}{c}{\underline{$m_{\tilde f}$ [TeV]}}\\
${\tilde u}_L,{\tilde c}_L$& 2.42 &\graycell 3.75 &${\tilde u}_R,{\tilde c}_R$& 4.11 &\graycell 4.72 &${\tilde t}_1$& 0.695 &\graycell 1.29 &${\tilde t}_2$& 3.16 &\graycell 1.81 \\
${\tilde d}_L,{\tilde s}_L$& 2.42 &\graycell 3.75 &${\tilde d}_R,{\tilde s}_R$& 4.7 &\graycell 7.22 &${\tilde b}_1 \approx{\tilde b}_L$& 0.689 &\graycell 1.61 &${\tilde b}_2 \approx{\tilde b}_R$& 4.28 &\graycell 7 \\
${\tilde \nu^e}_L,{\tilde \nu^\mu}_L$& 4.65 &\graycell 5.74 &${\tilde \nu^e}_R,{\tilde \nu^\mu}_R$& 3.05 &\graycell 4.95 &${\tilde \nu^\tau}_L$& 4.38 &\graycell 5.6 &${\tilde \nu^\tau}_R$& 3.05 &\graycell 4.95 \\
${\tilde e}_L,{\tilde \mu}_L$& 4.65 &\graycell 5.74 &${\tilde e}_R,{\tilde \mu}_R$& 12.2 &\graycell 18.1 &${\tilde \tau}_1\approx {\tilde \tau}_L$& 4.38 &\graycell 5.6 &${\tilde \tau}_2\approx {\tilde \tau}_R$& 12 &\graycell 18 \\
\cbottomrule
\end{tabular}
}
\caption{Sfermion masses for point 1 (white background) and
  point 2 (gray background).  The mass mixing for third family
    sfermions is small except for the stops in the second point where
    the two diagonal entries in \refeq{eq:mass} are similar resulting
    in a large mixing. The values for the first (second) illustration
    point are $\sin \theta_{{\tilde t}_L {\tilde t}_R}=0.04(0.65)$,
    $\sin \theta_{{\tilde b}_L {\tilde b}_R}=0.007(0.003)$ and $\sin
    \theta_{{\tilde \tau}_L {\tilde \tau}_R}=0.0006(0.0003)$.}
\label{tab:MSSM_sfermions}
\end{center}
\end{table}

\begin{figure}[!ht]
\begin{center}
\includegraphics[width=7.75cm]{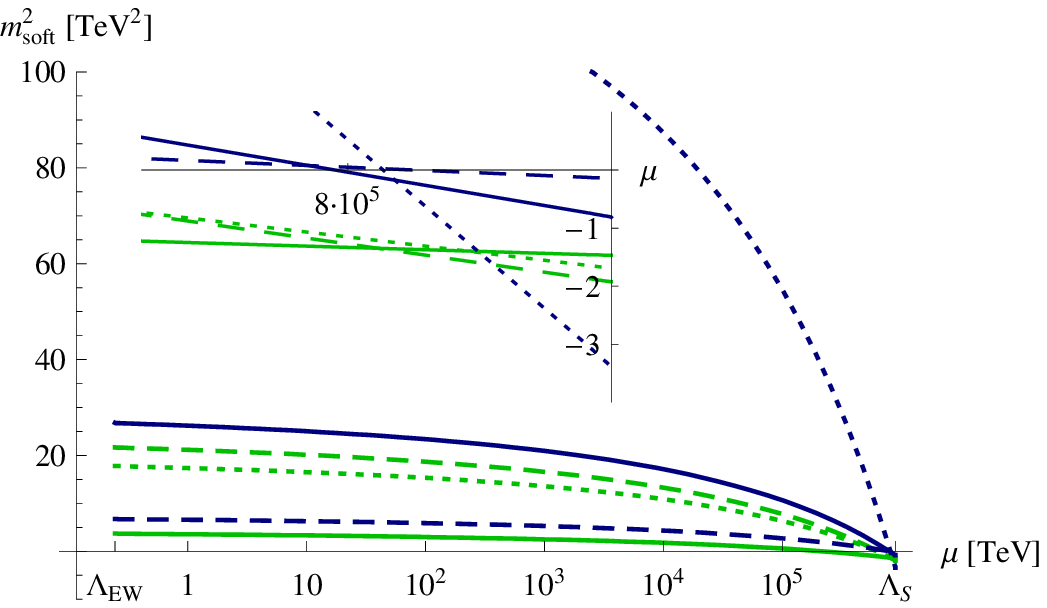}
\includegraphics[height=5.2cm,width=1cm]{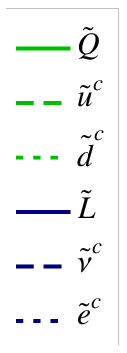}
\includegraphics[width=7.75cm]{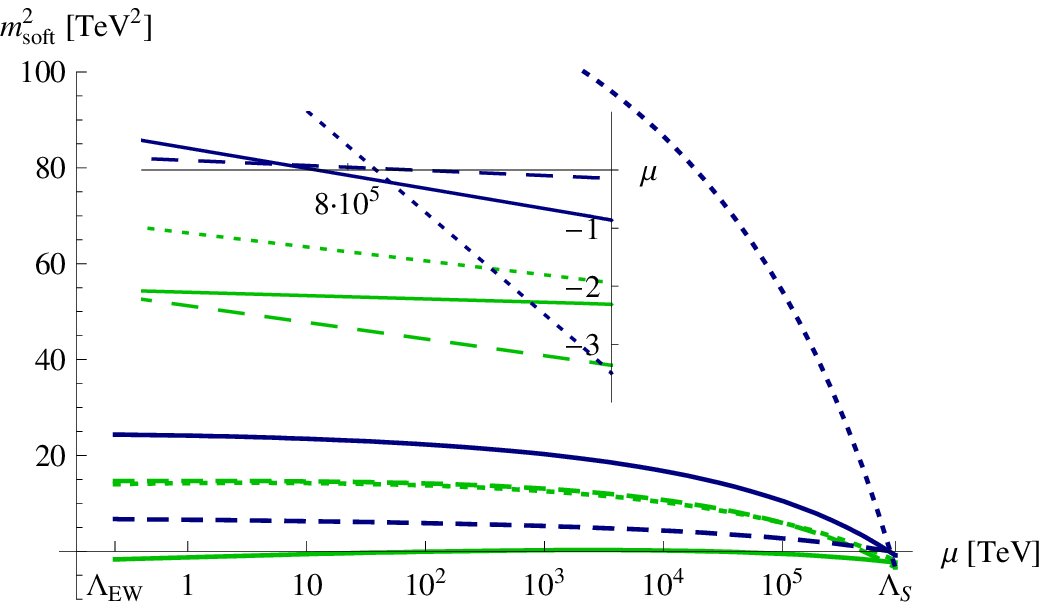}
\caption{Running of the MSSM sfermion soft masses squared in the first
illustration point. First and second families (left). Third family
(right). A close view of the region near $\Lambda_S$ is displayed to
illustrate the pure anomaly mediation contribution.
\label{Sleptons}}
\label{fig:sfermion_masses}
\end{center}
\end{figure}

The MSSM sfermions are in the range of a few TeV. They range from the
lightest, which is the lighter sbottom (stop) for the first (second)
point, to the heaviest, which is the right-handed slepton. Their
masses are given in Table \ref{tab:MSSM_sfermions}. The MSSM sfermion
masses are determined by the contributions from anomaly mediation,
$\zp$-gaugino mediation, and D-term contributions after $U(1)'$ gauge
symmetry breaking. Therefore, most features of the spectrum can be
understood from the prediction of pure anomaly mediation and the
choices of charges in Table~\ref{tab:charges}. The anomaly
contribution to the soft masses is negative not only for the first and
second families of sleptons but for all the MSSM sfermions. This is
because of the vanishing of the strong coupling $\beta$-function at
one loop and the extra negative contribution from the $\beta$-function
of the $\zp$ gauge coupling. The $\zp$-gaugino mediation leads to
large positive contributions proportional to $Q_i^2$, raising the soft
masses from the tachyonic region, as can be seen in
Fig.~\ref{fig:sfermion_masses}. This dependence on the charge explains
the difference between the first two generations of left-handed and
right-handed squarks. The third generation squarks are generically
lighter due to the effect of the Yukawa couplings in the running,
which can in some cases turn some of the soft masses tachyonic
again. From Fig.~\ref{fig:sfermion_masses} (right) we observe that
this is the case for the left-handed stop and sbottom in our
particular example. The $D$-term contributions can be either positive
or negative depending on $Q_i Q_S$. Although in general smaller, here
they are responsible for returning the squared masses for the
left-handed squarks back to the positive region.

%
%

%

Similarly, the exotic sfermions range from the lightest,
$\tilde{D}_1$, to the heaviest, $\tilde{E}_2$, which is also the
heaviest sfermion in the spectrum. Their masses and mixings are given
in Table \ref{tab:exotic_sfermions}. The exotic fermion masses are
given in Table \ref{tab:exotic_fermions}.
%
\begin{table}[!ht]
\begin{center}
{\small
\begin{tabular}{c c c | c c c | c c }
\ctoprule
&\multicolumn{2}{c}{$m_{\tilde f}$ [TeV]}\vline&&\multicolumn{2}{c}{$m_{\tilde f}$ [TeV]}\vline&\multicolumn{2}{c}{$\sin \theta_{{\tilde f}{\tilde f}^c}$}\\
\midrule
${\tilde D}_1$& 2.53 &\graycell 3.24 &${\tilde D}_2$& 6.41 &\graycell 11.6 & 0.48 &\graycell 0.63 \\
${\tilde E}_1$& 9.25 &\graycell 15.6 &${\tilde E}_2$& 12.8 &\graycell 20.6 & 0.37 &\graycell 0.55 \\
\cbottomrule
\end{tabular}
}
\caption{Exotic sfermion masses and mixings for point 1 (white
  background) and point 2 (gray background).}
\label{tab:exotic_sfermions}
\end{center}
\end{table}

\begin{table}[!ht]
\begin{center}
{\small
\begin{tabular}{c c c }
\ctoprule
&\multicolumn{2}{c}{$m_{f}$ [TeV]}\\
\midrule
$D$& 3.57 &\graycell 7.56 \\
$E$& 5.95 &\graycell 12.6 \\
\cbottomrule
\end{tabular}
}
\caption{Exotic fermion masses for point 1 (white background) and
  point 2 (gray background).}
\label{tab:exotic_fermions}
\end{center}
\end{table}

\begin{figure}[!ht]
\begin{center}
\includegraphics[width=\textwidth]{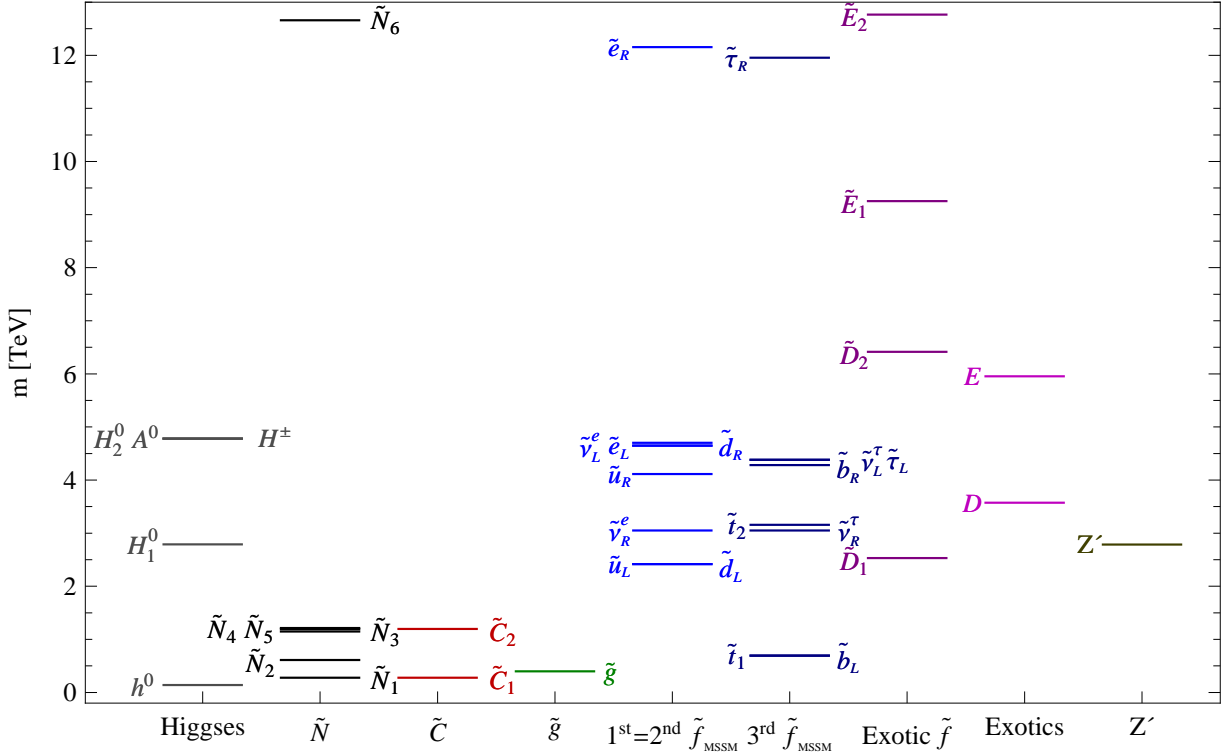}
\caption{Spectrum of the first illustration point discussed in the text.}
\label{fig:spectrum1}
\end{center}
\end{figure}

\begin{figure}[!ht]
\begin{center}
\includegraphics[width=\textwidth]{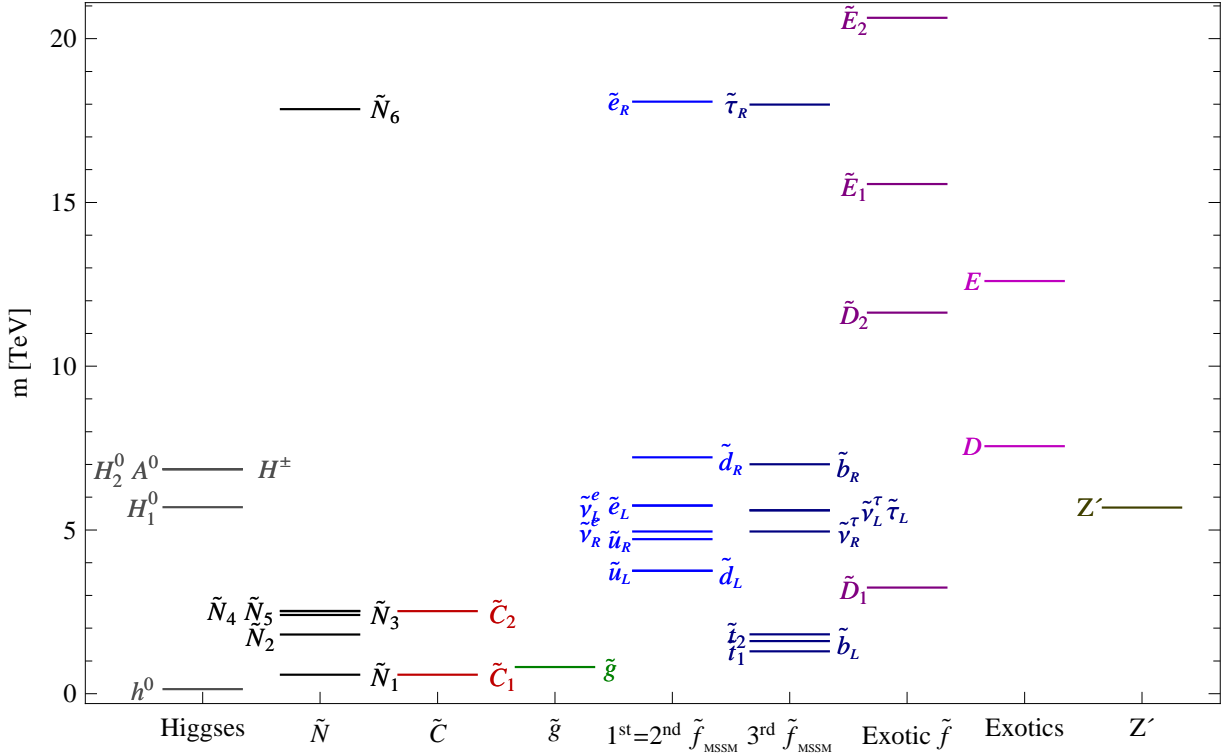}
\caption{Spectrum of the second illustration point discussed in the text.}
\label{fig:spectrum2}
\end{center}
\end{figure}

Finally, the $\zp$ gauge boson mass and $Z-\zp$ mixing angle are
listed in Table \ref{tab:Zp_boson}. The spectra are summarized in
Figs.~\ref{fig:spectrum1} and \ref{fig:spectrum2}.

\begin{table}[!ht]
\begin{center}
{\small
\begin{tabular}{c c | c c }
\ctoprule
\multicolumn{2}{c}{$M_{\zp}$ [TeV]}\vline&\multicolumn{2}{c}{$\sin \theta_{Z\zp}$}\\
\midrule
 ~~2.78~~ &\graycell ~~5.68~~ & ~3$\cdot$10$^{-4}$ &\graycell ~7$\cdot$10$^{-5}$ \\
\cbottomrule
\end{tabular}
}
\caption{$\zp$ mass and mixing for point 1 (white background) and
  point 2 (gray background). }
\label{tab:Zp_boson}
\end{center}
\end{table}

%
%

%
%

The fact that $m_{H_1}$ and $M_{Z'}$ are approximately the same is not
an accident. It arises from the fact that there is very little mixing
in the Higgs sector. Consequently $H_1$ is almost ``pure" $S$. If we
now consider only the $S$ and $\zp$ sector and allow for negative
$m_S^2$, we can think of $m_S^2$ as arising from a Fayet-Iliopoulos
term \cite{Fayet:1974jb}. Adding a constant term to the scalar
potential for $S$, we can write it as,
\begin{equation}
V=m^2_S|S|^2+\frac{\gz^2}{2} Q_S^2 |S|^4+\mbox{constant}=
\frac{1}{2}\left(\xi-\gz Q_S|S|^2\right)^2,
\end{equation}
where $\xi=-m_S^2/\gz Q_S$. The vacuum in this case breaks the $U(1)'$
gauge symmetry but not supersymmetry. The chiral and the vector
multiplet are combined to form a massive vector multiplet\footnote{The
  singlino and the $\zp$ gaugino get their mass from the term
  $-\sqrt{2}\gz Q_s S \tilde S \tilde{Z}^\prime$ +h.c. in the
  Lagrangian.} with mass $\sqrt{2} |m_S|$.

Introducing a gaugino mass term breaks supersymmetry explicitly, but
at tree level its presence only shifts the singlino and the $\zp$
gaugino mass, while the scalar and vector components of the
supermultiplet remain degenerate. This degeneracy will be lifted by
other interactions of the scalar component of $S$, and its mixing, but
these are still smaller effects, so to a good approximation $m_{H_1}$
and $M_{Z'}$ are equal.

\section{Conclusions}\label{sec:conclusions}

In this paper, we argued that combining $\zp$ mediation with anomaly
mediation is both a plausible and a feasible scenario. As an example ,
we have considered a particular realization and argued that its
contribution can be naturally comparable to anomaly mediation. Such an
approach solves the tachyonic slepton mass problem of anomaly
mediation, and the need for fine-tuning in the original $\zp$
mediation models. In this context, it is natural to consider an
NMSSM-like model in which $U(1)'$ gauge symmetry forbids a $\mu$ term
in the superpotential. In this case, the $U(1)'$ gauge symmetry
breaking, the effective $\mu$ and $B_\mu$ terms, and the electroweak
symmetry breaking must all be generated dynamically from UV input
values. We found that it is not difficult to find viable models in
this scenario. We presented two explicit examples with different low
energy spectrum.

We comment on generic phenomenological features of this class of
models. The gaugino spectrum is completely determined by the anomaly
mediation. However, since we have to introduce additional exotic
matter, the anomaly mediation contribution can be dramatically
different from the MSSM prediction. Such a change can lead to
significantly different phenomenology. In the specific class of models
we considered here, the gluino only receives a contribution from
2-loop anomaly mediation, and is only somewhat heavier than the LSP
(wino in this case). Such light gluinos can be copiously produced at
the LHC. Since stops and sbottoms are typically lighter than the first
two families of squarks, the decay products of the gluino are mainly
dominated by the third generation states: $b \bar{b} + \Tilde{N}_0 $,
$t + \bar{b} + \Tilde{C}^{+}$, and $t \bar{t} + \Tilde{N}_0$. However,
the availability of decay channels involving top quarks depends on
$M_{3} - M_{2}$. For the first illustration point $M_{3} - M_{2} <
m_t$ and only the $b \bar{b} + \Tilde{N}_0 $ decay mode is
possible. For the second illustration point $m_t< M_{3} - M_{2} <2m_t$
and both $b \bar{b} + \Tilde{N}_0 $ and $t + \bar{b} + \Tilde{C}^{+}$
are possible. Requiring the gluino is lighter than a TeV, the largest
value of $M_{3} - M_{2}$ is of the order of $300$ GeV. Different
choices of the input charges may help in getting an even larger
splitting. While the $b \bar{b} b \bar{b} + \not{\!\!E}_T$ signal is a
very useful discovery channel, the decay channels which lead to
multiple top final states have spectacular signals and can lead to
early discovery at the LHC \cite{Acharya:2009gb}. Another prominent
feature of this scenario is the presence of a $\zp$ gauge boson with
$M_{\zp} $ around several TeV, unlike the pure $\zp$-gaugino mediation
where $M_{\zp}$ is typically very heavy. Such a $\zp$ has an excellent
chance of being discovered at the LHC. Detailed measurements of its
properties, in particular its couplings to various Standard Model
matter fields, especially leptons and third generation quarks, provide
clues crucial to piecing together the complete picture of mediation of
supersymmetry breaking. Such a $\zp$ will also decay into
superpartners, which also offers a good opportunity of studying their
properties \cite{Baumgart:2006pa,Han:2009ss}. In particular, in this
model, $\zp$ decay probably offers the only possibility of discovering
and study the properties of the singlino.

\vspace{0.3cm}

{\em Acknowledgments:} We would like to thank Zohar Komargodski, Yael
Shadmi and Jay Wacker for useful discussions. J.B. also gratefully
acknowledges the hospitality of the Institute for Advanced Study at
Princeton during part of this work.  The work of J.B. is supported by
MICINN project FPA2006-05294 and Junta de Andaluc\'ia projects FQM
101, FQM 437 and FQM 03048. The work of P.L. is supported by the IBM
Einstein Fellowship and by NSF grant PHY-0503584.  The work of G.P. is
supported in part by the Department of Energy grants DE-FG02-90ER40542
and DE-FG02-90ER40560, and by the United States-Israel Bi-national
Science Foundation grant 2006280. The work of L.-T. W. is supported by
the National Science Foundation under grant PHY-0756966 and the
Department of Energy under grant DE-FG02-90ER40542.

\begin{appendix}

\section{\boldmath $U(1)^\prime$ charges}\label{app:charges}
We normalize all the $U(1)'$ charges such that $Q_{H_d}=1$. Defining
$Q_{H_u}=x$ and $Q_{Q}=y$, the other charges are
\begin{eqnarray}
&&Q_{u^c}=-x-y,\quad Q_{d^c}=-1-y,\quad Q_L=-\frac13(1+x)-3y,\nonumber\\
&&Q_{e^c}=-1+\frac{1+x}{3}+3y,\quad Q_{\nu^c}=-x+\frac{1+x}{3}+3y\nonumber\\
&&Q_S=-1-x,\quad Q_D=\frac29 (3 + x - 3 y) + \sqrt{2(1 + 3 y) (1 - x + 6 y)}, \quad Q_{D^c}=-Q_S-Q_D\nonumber\\
&&Q_E=-3y-\frac32Q_D+2,\quad Q_{E^c}=-Q_S-Q_E.
\end{eqnarray}

\section{Boundary conditions}\label{app:bc}
We assume that at a scale $\Lambda_S$ the gaugino and scalar masses
are generated from the anomaly contribution.  We use the general
expressions from \cite{Gherghetta:1999sw},
\begin{eqnarray}
M_\lambda &=&
-\frac{g^2}{2}\frac{d g^{-2}}{d \ln \mu} m_{3/2}
=
\frac{\beta_g}{g} m_{3/2}
\nonumber
\\
  m_{\tilde Q}^2&=&-\frac14
\frac{d^2 \ln Z_Q}{d(\ln \mu)^2} m_{3/2}^2=
-\frac14
      \left(\frac{\partial\gamma}{\partial g}\beta_g +
       \frac{\partial\gamma}{\partial y}\beta_y\right)m_{3/2}^2 \nonumber\\
   A_{y}&=&\frac{1}{2}
    \sum_i \frac{d\ln Z_{Q_i}}{d\ln \mu} m_{3/2}=
        -{\beta_y} m_{3/2},
\end{eqnarray}
where $\gamma\equiv {d \ln Z}/{d\ln\mu}$, $\beta_g\equiv {d
g}/{d\ln\mu}$, $\beta_y\equiv {d y}/{d\ln\mu}$, and $A_y$ is defined
as in ${\cal L}=-A_y~ \phi_1\,\phi_2\,\phi_3$ +h.c.. At one loop
\begin{equation}
\gamma^i_j=-\frac{1}{16\pi^2}\left[y^{imn}y^*_{jmn}-4g_a^2C_a(i)\delta^i_j\right].
\end{equation}

To determine the boundary conditions, we need the beta functions for
the gauge and Yukawa couplings. In the mixed scenario the dominant
contribution to the gaugino masses (apart from the $Z'$ gaugino) is
the anomaly contribution.  For the models of \cite{Langacker:2007ac,
Langacker:2008ip}, the beta function for $g_3$ vanishes at one
loop. As a result at one loop the gluino is massless.  It will get a
non-zero contribution at two-loop order. For consistency we will use
the two-loop expressions for all three gauge coupling beta
functions. To derive the beta functions we use the general expressions
in \cite{Martin:1993zk,Yamada:1993ga,Yamada:1994id, Jack:1994kd}. In the following
$n_D=3$ and $n_E=2$.

\subsection{Gauge and Yukawa $\beta$ functions}
The gauge coupling  $\beta$ functions are
\begin{eqnarray}\label{eq:gauge_beta}
\beta_{g_1}&=&\frac{g_1^3}{16\pi^2}\bigg\{\frac{51}{5}+\frac{1}{16\pi^2}
\bigg[24g_3^2+\frac{27}{5}g_2^2+\frac{351}{25} g_1^2+
g_{z'}^2 \frac{12}{5}{\rm Tr}\left(Y^2Q^2\right)\nonumber\\
&&\qquad\quad-\frac{26}{5}y_t^2-\frac{14}{5}y_b^2-\frac{18}{5}y_\tau^2-\frac65\lambda^2
-\frac{12}5y_D^2-\frac{24}5y_E^2\bigg]\bigg\}\nonumber\\
\beta_{g_2}&=&\frac{g_2^3}{16\pi^2}\bigg\{1+\frac{1}{16\pi^2}\bigg[24g_3^2+25g_2^2+\frac95
  g_1^2+2g_{z'}^2 (Q_{H_d}^2 + Q^2_{H_u} + 3 (Q_L^2 + 3Q_Q^2))\nonumber\\
&&\qquad\quad -6y_t^2-6y_d^2-2y_\tau^2-2\lambda^2\bigg]\bigg\}\nonumber\\
\beta_{g_3}&=&\frac{g_3^3}{(16\pi^2)^2}\bigg[48g_3^2+9g_2^2+3g_1^2+6g_{z'}^2(Q_D^2  
+ Q_{D^c}^2 + 2 Q_Q^2 + Q_{u^c}^2+ Q_{d^c}^2)-4y_t^2-4y_b^2-6y_D^2\bigg]\nonumber\\
\beta_{\gz}&=&\frac{\gz^3}{16\pi^2}\bigg\{{\rm
Tr}\left(Q^2\right)+\frac{1}{16\pi^2}\bigg[4\gz^2{\rm
Tr}\left(Q^4\right)+
g_{1}^2 \frac{12}{5}{\rm Tr}\left(Y^2Q^2\right)\nonumber\\
&&\qquad\quad +6g_{2}^2 (Q_{H_d}^2 + Q^2_{H_u} + 3 (Q_L^2 + 3Q_Q^2))
+48g_{3}^2(Q_D^2  + Q_{D^c}^2 + 2 Q_Q^2 + Q_{u^c}^2+ Q_{d^c}^2)-
\nonumber\\
&&\qquad\quad-12(Q_{u^c}^2 + Q_Q^2 + Q_{H_u}^2)y_t^2-12(Q_{d^c}^2 + Q_Q^2 + Q_{H_d}^2)y_b^2 
-4(Q_{e^c}^2 + Q_L^2 + Q_{H_d}^2)y_e^2\nonumber\\
&&\qquad\quad-18(Q_{D^c}^2 + Q_D^2 + Q_S^2)y_D^2-
4(Q_{E^c}^2 + Q_E^2 + Q_S^2)y_E^2
-4(Q_{H_u}^2 + Q_{H_d}^2 + Q_S^2)\lambda^2\bigg]\bigg\}.\nonumber\\
\end{eqnarray}
The relevant $\beta$ functions for the Yukawa couplings are
\begin{eqnarray}\label{eq:Yukawa_beta}
\beta_{y_t}&=&\frac{y_t}{16\pi^2}\bigg[\lambda^2+6y_t^2+y_b^2-\frac{16}{3}g_3^2-3g_2^2
-\frac{13}{15} g_1^2-2\gz^2(Q_{H_u}^2+Q_Q^2+Q_{u^c}^2)\bigg]\nonumber\\
\beta_{y_b}&=&\frac{y_b}{16\pi^2}\bigg[\lambda^2+6y_b^2+y_t^2+y_e^2-\frac{16}{3}g_3^2-3g_2^2
-\frac{7}{15} g_1^2-2\gz^2(Q_{H_d}^2+Q_Q^2+Q_{d^c}^2)\bigg]\nonumber\\
\beta_{y_\tau}&=&\frac{y_\tau}{16\pi^2}\bigg[\lambda^2+3y_b^2+4y_\tau^2-3g_2^2
-\frac{9}{5} g_1^2-2\gz^2(Q_{H_d}^2+Q_L^2+Q_{e^c}^2)\bigg]\nonumber\\
\beta_\lambda&=&\frac{\lambda}{16\pi^2}\bigg[4\lambda^2+3y_t^2+3y_b^2+y_\tau^2+3n_D\,y_D^2+n_E\,y_E^2-3g_2^2
-\frac35 g_1^2-2g^2_{z'}\left(Q_S^2+Q_{H_u}^2+Q^2_{H_d}\right)\bigg] \nonumber\\
\beta_{y_D}&=&\frac{y_D}{16\pi^2}\bigg[2\lambda^2+(3n_D+2)y_D^2+n_Ey_E^2-\frac{16}{3}g_3^2
-\frac65 g_1^2(Y_D^2+Y_{D^c}^2)-2\gz^2(Q_S^2+Q_D^2+Q_{D^c}^2)\bigg]\nonumber\\
\beta_{y_E}&=&\frac{y_E}{16\pi^2}\bigg[2\lambda^2+3n_Dy_D^2+(n_E+2)y_E^2-
\frac65 g_1^2(Y_E^2+Y_{E^c}^2)-2\gz^2(Q_S^2+Q_E^2+Q_{E^c}^2)\bigg].\nonumber\\
\end{eqnarray}
We have set all the SM Yukawas, apart from $y_t,\,y_b$, and $y_\tau$,  to zero.
\subsection{Gaugino masses}
The MSSM gaugino masses are
\begin{eqnarray}
M_3(\Lambda_S)=\frac{\beta_{g_3}}{g_3}\,m_{3/2}, \quad M_2(\Lambda_S)=
\frac{\beta_{g_2}}{g_2}\,m_{3/2}, \quad M_1(\Lambda_S)=\frac{\beta_{g_1}}{g_1}\,m_{3/2}.
\end{eqnarray}
The $Z'$ gaugino mass, $\mzp$, is a free parameter. If we fix it at the scale $\mzp$, its value at $\Lambda_S$ would be
\begin{equation}
\mzp(\Lambda_S)=\mzp(\mzp)\left[1-\frac{{\rm Tr}\left(Q^2\right)\gz^2(\Lambda_s)}
{8\pi^2}\ln\left(\frac{\Lambda_S}{\mzp}\right)\right].
\end{equation}
\subsection{Scalar masses}\label{sec:scalar_bc}
The general expression for the scalar masses is schematically
\begin{equation}
m^2=\frac{m_{3/2}^2}{16\pi^2}\left[\#\,y\,\beta_y-\sum_i\,2\,C_i\,g_i\,\beta_i\right],
\end{equation}
where $\#$ is an integer which depends on the specific form of the Yukawa coupling. The constants $C_i$ are
\begin{equation}
C_3=4/3,\quad C_2=3/4,\quad C_1=3\,Y^2/5,\quad C_{z'}=Q^2,
\end{equation}
where $Y$ is the hypercharge and $Q$ is the $U(1)'$ charge.

The expression for the soft masses at the SUSY breaking scale are, for
$S,H_u,$ and $H_d$,
\begin{eqnarray}
m^2_S&=&\frac{m^2_{3/2}}{16\pi^2}\left(2\lambda\,\beta_\lambda+3n_Dy_D\,
\beta_{y_D}+n_Ey_E\,\beta_{y_E}-2\gz Q_S^2\,\beta_{\gz}\right)\nonumber\\
m^2_{H_u}&=&\frac{m^2_{3/2}}{16\pi^2}\left(\lambda\,\beta_\lambda+3y_t\,
\beta_{y_t}-\frac64 g_2\,\beta_{g_2}-\frac65Y^2_{H_u}g_1\,\beta_{g_1}
-2\gz Q_{H_u}^2\,\beta_{\gz}\right)\nonumber\\
m^2_{H_d}&=&\frac{m^2_{3/2}}{16\pi^2}\left(\lambda\,\beta_\lambda
+3y_b\,\beta_{y_b}+y_\tau\,\beta_{y_\tau}-\frac64 g_2\,\beta_{g_2}-\frac65Y^2_{H_d}g_1\,\beta_{g_1}
-2\gz Q_{H_d}^2\,\beta_{\gz}\right);
\end{eqnarray}
for the scalar exotics,
\begin{eqnarray}
m^2_{\tilde D}&=&\frac{m^2_{3/2}}{16\pi^2}\left(y_D\,\beta_{y_D}-\frac83 g_3\,
\beta_{g_3}-\frac65Y^2_{D}g_1\,\beta_{g_1}
-2\gz Q_{D}^2\,\beta_{\gz}\right)\nonumber\\
m^2_{\tilde{D}^c}&=&\frac{m^2_{3/2}}{16\pi^2}\left(y_{D}\,\beta_{y_D}-\frac83 g_3\,
\beta_{g_3}-\frac65Y^2_{D^c}g_1\,\beta_{g_1}
-2\gz Q_{D^c}^2\,\beta_{\gz}\right)\nonumber\\
m^2_{\tilde E}&=&\frac{m^2_{3/2}}{16\pi^2}\left(y_E\,\beta_{y_E}-\frac65Y^2_{E}g_1\,\beta_{g_1}
-2\gz Q_{E}^2\,\beta_{\gz}\right)\nonumber\\
m^2_{\tilde{E}^c}&=&\frac{m^2_{3/2}}{16\pi^2}\left(y_{E}\,\beta_{y_E}-\frac65Y^2_{E^c}g_1\,\beta_{g_1}
-2\gz Q_{E^c}^2\,\beta_{\gz}\right);
\end{eqnarray}
for the third generation squarks
\begin{eqnarray}
m^2_{\tilde Q_3}&=&\frac{m^2_{3/2}}{16\pi^2}\left(y_t\,\beta_{y_t}+y_b\,\beta_{y_b}
-\frac83 g_3\,\beta_{g_3}-\frac64 g_2\,\beta_{g_2}-\frac65Y^2_{Q}g_1\,\beta_{g_1}
-2\gz Q_{Q}^2\,\beta_{\gz}\right)\nonumber\\
m^2_{\tilde{t}^c}&=&\frac{m^2_{3/2}}{16\pi^2}\left(2y_t\,\beta_{y_t}-\frac83 g_3\,\beta_{g_3}
-\frac65Y^2_{u^c}g_1\,\beta_{g_1}
-2\gz Q_{u^c}^2\,\beta_{\gz}\right)\nonumber\\
m^2_{\tilde{b}^c}&=&\frac{m^2_{3/2}}{16\pi^2}\left(2y_b\,\beta_{y_b}
-\frac83 g_3\,\beta_{g_3}-\frac65Y^2_{d^c}g_1\,\beta_{g_1}
-2\gz Q_{d^c}^2\,\beta_{\gz}\right);
\end{eqnarray}
for the first two generations of squarks,
\begin{eqnarray}
m^2_{\tilde{Q}_i}&=&\frac{m^2_{3/2}}{16\pi^2}\left(-\frac83 g_3\,\beta_{g_3}
-\frac64 g_2\,\beta_{g_2}-\frac65Y^2_{Q}g_1\,\beta_{g_1}
-2\gz Q_{Q}^2\,\beta_{\gz}\right)\nonumber\\
m^2_{\tilde{u}^c_i}&=&\frac{m^2_{3/2}}{16\pi^2}\left(-\frac83 g_3\,\beta_{g_3}-\frac65 Y^2_{u^c}g_1\,\beta_{g_1}
-2\gz Q_{u^c}^2\,\beta_{\gz}\right)\nonumber\\
m^2_{\tilde{d}^c_i}&=&\frac{m^2_{3/2}}{16\pi^2}\left(-\frac83 g_3\,\beta_{g_3}-\frac65Y^2_{d^c}g_1\,\beta_{g_1}
-2\gz Q_{d^c}^2\,\beta_{\gz}\right);
\end{eqnarray}
for the third generation of charged sleptons,
\begin{eqnarray}
m^2_{\tilde{L}_3}&=&\frac{m^2_{3/2}}{16\pi^2}\left(y_\tau\,\beta_{y_\tau}-\frac64 g_2\,\beta_{g_2}-\frac65Y^2_{L}g_1\,\beta_{g_1}
-2\gz Q_{L}^2\,\beta_{\gz}\right)\nonumber\\
m^2_{\tilde{\tau }^c}&=&\frac{m^2_{3/2}}{16\pi^2}\left(2y_\tau\,\beta_{y_\tau}-\frac65Y^2_{e^c}g_1\,\beta_{g_1}
-2\gz Q_{e^c}^2\,\beta_{\gz}\right);
\end{eqnarray}
and for the rest of the sleptons,
\begin{eqnarray}
m^2_{\tilde{L}_i}&=&\frac{m^2_{3/2}}{16\pi^2}\left(-\frac64 g_2\,\beta_{g_2}-\frac65Y^2_{L}g_1\,\beta_{g_1}
-2\gz Q_{L}^2\,\beta_{\gz}\right)\nonumber\\
m^2_{\tilde{e}^c_i}&=&\frac{m^2_{3/2}}{16\pi^2}\left(-\frac65Y^2_{e^c}g_1\,\beta_{g_1}
-2\gz Q_{e^c}^2\,\beta_{\gz}\right)\nonumber\\
m^2_{\tilde{\nu}^c_i}&=&\frac{m^2_{3/2}}{16\pi^2}\left(-2\gz Q_{\nu^c}^2\,\beta_{\gz}\right).
\end{eqnarray}

\subsection{$A$ terms}
The non-zero Yukawa couplings are $y_t,y_b,y_\tau,\,\lambda,\,y_D$ and
$y_E$. The corresponding $A$ terms are:
\begin{eqnarray}
&&A_{t}=-\beta_{y_t}m_{3/2},\quad
A_{b}=-\beta_{y_b}m_{3/2},\quad
A_{\tau}=-\beta_{y_\tau}m_{3/2},\quad\nonumber\\
&&A_\lambda=-\beta_{\lambda}m_{3/2},\quad
A_{D}=-\beta_{y_D}m_{3/2},\quad A_{E}=-\beta_{y_E}m_{3/2}.
\end{eqnarray}

\section{RGE equations}\label{app:RGEs}
To derive the RGEs we use the general expressions in
\cite{Martin:1993zk,Yamada:1993ga,Yamada:1994id,Jack:1994kd}.
\subsection{Gauge and Yukawa couplings}
The gauge coupling RGEs are
\begin{equation}
d g_i/d\ln\mu=\beta_{g_i},
\end{equation}
where $i=1,2,3,Z^\prime$ and the $\beta_i$ are given in (\ref{eq:gauge_beta}).

The Yukawa RGEs are
\begin{equation}
d y_i/d\ln\mu=\beta_{y_i},
\end{equation}
where $i=t,b,\tau,\lambda,D,E$ and the $\beta_i$ are given in (\ref{eq:Yukawa_beta}).

\subsection{Gaugino masses}
The RGEs for the gaugino masses are
\begin{eqnarray}
\frac{dM_1}{d\ln\mu}&=&\frac{g_1^2}{16\pi^2}\Bigg\{\frac{102}{5}M_1+
\frac{2}{16\pi^2}\bigg[24g_3^2(M_1+M_3)+\frac{27}{5}g_2^2(M_1+M_2)+\frac{351}{25}
g_1^2(2M_1)\nonumber\\
&& + g_{z'}^2 \frac{12}{5}{\rm Tr}
\left(Y^2Q^2\right)(M_1+\mzp))\nonumber\\
&& +M_1\Big(
-\frac{26}{5}y_t^2-\frac{14}{5}y_b^2-\frac{18}{5}y_\tau^2-\frac65\lambda^2
-\frac{12}5y_D^2-\frac{24}5y_E^2\Big)\nonumber\\
&&+\frac{26}{5}A_t y_t+\frac{14}{5}A_b y_b+\frac{18}{5}A_\tau y_\tau+\frac65
A_\lambda\lambda +\frac{12}5A_D y_D+\frac{24}5 A_Ey_E\bigg]\Bigg\}\nonumber\\
\frac{dM_2}{d\ln\mu}&=&\frac{g_2^2}{16\pi^2}\Bigg\{2M_2+\frac{2}{16\pi^2}\bigg[24g_3^2(M_2+M_3)+25g_2^2(2M_2)+\frac95
g_1^2(M_2+M_1)\nonumber\\
&& +2g_{z'}^2 (Q_{H_d}^2 + Q^2_{H_u} + 3 (Q_L^2 +
3Q_Q^2))(M_2+\mzp)\nonumber\\
&&+M_2\Big(-6y_t^2-6y_b^2-2y_\tau^2-2\lambda^2\Big)
+6A_ty_t+6A_by_b+2A_\tau y_\tau+2A_\lambda\lambda\bigg]\Bigg\}\nonumber\\
\frac{dM_3}{d\ln\mu}&=&\frac{2g_3^2}{(16\pi^2)^2}\bigg[48g_3^2(2M_3)+
9g_2^2(M_3+M_2)+3g_1^2(M_3+M_1)+M_3\Big(-4y_t^2-4y_b^2-6y_D^2\Big)\nonumber\\
&&+6g_{z'}^2(Q_D^2 + Q_{D^c}^2 +
2 Q_Q^2 + Q_{u^c}^2+ Q_{d^c}^2)(M_3+\mzp)+4A_ty_t+4A_by_b+6A_Dy_D\bigg]\nonumber\\
\frac{d\mzp}{d\ln\mu}&=&\frac{\gz^2}{16\pi^2}\Bigg\{2{\rm
Tr}\left(Q^2\right)\mzp+\frac{2}{16\pi^2}\bigg[4\gz^2{\rm
Tr}\left(Q^4\right)(2\mzp)+
g_{1}^2 \frac{12}{5}{\rm Tr}\left(Y^2Q^2\right)(\mzp+M_1)\nonumber\\
&&+6g_{2}^2 (Q_{H_d}^2 + Q^2_{H_u} + 3 (Q_L^2 + 3Q_Q^2))(\mzp+M_2)\nonumber\\
&&+48g_{3}^2(Q_D^2  + Q_{D^c}^2 + 2 Q_Q^2 + Q_{u^c}^2+
Q_{d^c}^2)(\mzp+M_3)+\nonumber\\
&&+
12(Q_{u^c}^2 + Q_Q^2 + Q_{H_u}^2)(A_t-\mzp y_t) y_t
+12(Q_{d^c}^2 + Q_Q^2 + Q_{H_d}^2)(A_b-\mzp y_b) y_b+\nonumber\\
&&+4(Q_{H_d}^2 +Q_L^2+Q_{e^c}^2)(A_\tau-\mzp y_\tau)+4(Q_{H_u}^2 + Q_{H_d}^2 +
Q_S^2)(A_\lambda-\mzp\lambda)+\nonumber\\
&&+18(Q_{D^c}^2 + Q_D^2 + Q_S^2)(A_D-\mzp y_D)y_D+ 4(Q_{E^c}^2 +
Q_E^2 + Q_S^2)(A_E-\mzp y_E)y_E \lambda\bigg]\Bigg\}.\nonumber\\
\end{eqnarray}

\subsection{Scalar masses}
The RGEs for the soft masses are given below. The $U(1)_Y$ and
$\uonep$ $D$-term contributions which are of the form Tr($Y m_i^2$)
and Tr($Q m_i^2$) are not included.  As explained in
\cite{Langacker:2008ip}, at one-loop order the RGEs for these traces
are homogeneous equations. Using the expressions of appendix
\ref{sec:scalar_bc}, one can show that these traces vanish at
$\mu=\Lambda_S$.  As a result they vanish for all scales and need not
be included in the RGEs for the soft masses. We have also verified
explicitly that numerically solving the soft masses RGEs with and
without these traces give the same result.

The expression for RGEs of the soft masses, are
for $S,H_u,$ and $H_d$,
\begin{eqnarray}
16\pi^2\,\frac{dm_S^2}{d\ln\mu}&=& - 8 \gz^2 Q_S^2 \mzp^2 +
4\lambda^2(m_S^2 + m_{H_u}^2 + m_{H_d}^2) \nonumber \\ &&+6 n_D
y_D^2(m_S^2 + m_{\tilde{D}}^2 + m_{\tilde{D}^c}^2) + 2 n_E y_E^2(m_S^2 + m_{\tilde E}^2 +
m_{\tilde{E}^c}^2)\nonumber \\ &&+4A^2_\lambda+2n_E A_E^2+6n_D
A_D^2\nonumber\\ 16\pi^2\,\frac{dm_{H_u}^2}{d\ln\mu}&=& - 8 \gz^2
Q_{H_u}^2 \mzp^2 -6M_2^2g_2^2-\frac{24}5M_1^2g_1^2Y_{H_u}^2
\nonumber\\ &&+ 2\lambda^2(m_S^2 + m_{H_u}^2 + m_{H_d}^2)+ 6y_t^2(
m_{H_u}^2 + m_{\tilde{Q}_3}^2+ m_{\tilde{t}^c}^2)+6A_t^2+ 2
A_\lambda^2\nonumber\\ 16\pi^2\,\frac{dm_{H_d}^2}{d\ln\mu}&=& - 8
\gz^2 Q_{H_d}^2 \mzp^2 -6M_2^2g_2^2-\frac{24}5M_1^2g_1^2Y_{H_d}^2
\nonumber\\ &&+ 2\lambda^2(m_S^2 + m_{H_u}^2 + m_{H_d}^2)+ 6y_b^2(
m_{H_d}^2 + m_{\tilde{Q}_3}^2+ m_{\tilde{b}^c}^2)+ 2y_\tau^2( m_{H_d}^2 + m_{\tilde{L}_3}^2+
m_{\tilde{\tau}^c}^2)\nonumber\\
&&+6A_b^2+2A_\tau^2+2 A_\lambda^2;
\end{eqnarray}
for the scalar exotics,
\begin{eqnarray}
16\pi^2\,\frac{dm_{\tilde D}^2}{d\ln\mu}&=& - 8 \gz^2 Q_D^2 \mzp^2
-\frac{32}{3}M_3^2g_3^2-\frac{24}5M_1^2g_1^2Y_D^2 +
2y_D^2(m_S^2 + m_{\tilde{D}}^2 + m_{\tilde{D}^c}^2)+ 2A_D^2\nonumber\\ 
16\pi^2\,\frac{dm_{\tilde{D}^c}^2}{d\ln\mu}&=& - 8
\gz^2 Q_{D^c}^2 \mzp^2
-\frac{32}{3}M_3^2g_3^2-\frac{24}5M_1^2g_1^2Y_{D^c}^2+
2y_D^2(m_S^2 + m_{\tilde{D}}^2 + m_{\tilde{D}^c}^2)+ 2
A_D^2\nonumber\\
16\pi^2\,\frac{dm_{\tilde E}^2}{d\ln\mu}&=& - 8 \gz^2
Q_E^2 \mzp^2 -\frac{24}5M_1^2g_1^2Y_E^2+ 2y_E^2(m_S^2 +
m_{\tilde{E}}^2 + m_{\tilde{E}^c}^2)+ 2
A_E^2\nonumber\\ 16\pi^2\,\frac{dm_{\tilde{E}^c}^2}{d\ln\mu}&=& - 8
\gz^2 Q_{E^c}^2 \mzp^2 -\frac{24}5M_1^2g_1^2Y_{E^c}^2 +
2y_E^2(m_S^2 + m_{\tilde E}^2 + m_{\tilde{E}^c}^2)+ 2 A_E^2;
\end{eqnarray}
for the third generation squarks,
\begin{eqnarray}
16\pi^2\,\frac{dm_{\tilde{Q}_3}^2}{d\ln\mu}&=& - 8 \gz^2 Q_{Q}^2
\mzp^2-\frac{32}{3}M_3^2g_3^2-6M_2^2g_2^2-\frac{24}5M_1^2g_1^2Y_Q^2\nonumber\\ 
&&+2y_t^2(m_{H_u}^2 + m_{\tilde{Q}_3}^2+
m_{\tilde{t}^c}^2)+2y_b^2(m_{H_d}^2 + m_{\tilde{Q}_3}^2+
m_{\tilde{b}^c}^2)+2A_t^2+2A_b^2\nonumber\\ 
16\pi^2\,\frac{dm_{\tilde{t}^c}^2}{d\ln\mu}&=&
- 8 \gz^2 Q_{\tilde{u}^c}^2
\mzp^2-\frac{32}{3}M_3^2g_3^2-\frac{24}5M_1^2g_1^2Y_{u^c}^2+4y_t^2(
m_{H_u}^2 + m_{\tilde{Q}_3}^2+
m_{\tilde{t}^c}^2)+4A_t^2\nonumber\\ 
16\pi^2\,\frac{dm_{\tilde{b}^c}^2}{d\ln\mu}&=&
- 8 \gz^2 Q_{d^c}^2
\mzp^2-\frac{32}{3}M_3^2g_3^2-\frac{24}5M_1^2g_1^2Y_{d^c}^2+4y_b^2(
m_{H_d}^2 + m_{\tilde{Q}_3}^2+
m_{\tilde{b}^c}^2)+4A_b^2;\nonumber\\ 
\end{eqnarray}
for the first two generations of squarks,
\begin{eqnarray}
16\pi^2\,\frac{dm_{\tilde{Q}_1}^2}{d\ln\mu}&=& - 8 \gz^2 Q_{Q}^2 \mzp^2-\frac{32}{3}M_3^2g_3^2-6M_2^2g_2^2-\frac{24}5M_1^2g_1^2Y_Q^2\nonumber\\
16\pi^2\,\frac{dm_{\tilde{u}_1^c}^2}{d\ln\mu}&=& - 8 \gz^2 Q_{u^c}^2 \mzp^2-\frac{32}{3}M_3^2g_3^2-\frac{24}5M_1^2g_1^2Y_{u^c}^2\nonumber\\
16\pi^2\,\frac{dm_{\tilde{d}_1^c}^2}{d\ln\mu}&=& - 8 \gz^2 Q_{d^c}^2 \mzp^2-\frac{32}{3}M_3^2g_3^2-\frac{24}5M_1^2g_1^2Y_{d^c}^2;
\end{eqnarray}
for the third generation of charged sleptons,
\begin{eqnarray}
16\pi^2\,\frac{dm_{\tilde{L}_3}^2}{d\ln\mu}&=& 
- 8 \gz^2 Q_{L}^2 \mzp^2-6M_2^2g_2^2-\frac{24}5M_1^2g_1^2Y_L^2+2y_\tau^2(
m_{H_d}^2 + m_{\tilde{L}_3}^2+
m_{\tilde{\tau}^c}^2)+2A_\tau^2\nonumber\\
16\pi^2\,\frac{dm_{\tilde{\tau}^c}^2}{d\ln\mu}&=& - 8 \gz^2 Q_{e^c}^2 \mzp^2-\frac{24}5M_1^2g_1^2Y_{e^c}^2+4y_\tau^2(
m_{H_d}^2 + m_{\tilde{L}_3}^2+
m_{\tilde{\tau}^c}^2)+4A_\tau^2;
\end{eqnarray}
and for the rest of the  sleptons,
\begin{eqnarray}
16\pi^2\,\frac{dm_{\tilde{L}}^2}{d\ln\mu}&=& - 8 \gz^2 Q_{L}^2 \mzp^2-6M_2^2g_2^2-\frac{24}5M_1^2g_1^2Y_L^2\nonumber\\
16\pi^2\,\frac{dm_{\tilde{e}^c}^2}{d\ln\mu}&=& - 8 \gz^2 Q_{e^c}^2 \mzp^2-\frac{24}5M_1^2g_1^2Y_{e^c}^2\nonumber\\
16\pi^2\,\frac{dm_{\tilde{\nu}^c}^2}{d\ln\mu}&=& - 8 \gz^2 Q_{\nu^c}^2 \mzp^2.
\end{eqnarray}

\subsection{$A$ terms}
The RGEs for the $A$ terms are
\begin{eqnarray}
16\pi^2\,\frac{dA_t}{d\ln\mu}&=&\frac{16}3g_3^2(2M_3y_t-A_t)+3g_2^2(2M_2y_t-A_t)+
\frac{13}{15}g_1^2(2M_1y_t-A_t)\nonumber\\
&&+2\gz^2(Q_{H_u}^2+Q_Q^2+Q_{u^c}^2)(2\mzp y_t-A_t)\nonumber\\
&&+18 A_ty_t^2+y_b^2 A_t+2 A_b y_b\, y_t+
\lambda^2 A_t+2 A_\lambda\lambda\, y_t\nonumber\\
16\pi^2\,\frac{dA_b}{d\ln\mu}&=&\frac{16}3g_3^2(2M_3y_b-A_b)+
3g_2^2(2M_2y_b-A_b)+\frac{7}{15}g_1^2(2M_1y_b-A_b)\nonumber\\
&&+2\gz^2(Q_{H_d}^2+Q_Q^2+Q_{d^c}^2)(2\mzp y_b-A_b)\nonumber\\
&&+18 A_by_b^2+y_t^2 A_b+
2 A_t y_t\, y_b+y_\tau^2 A_b+
2 A_\tau y_\tau\, y_b+\lambda^2 A_b+
2 A_\lambda\lambda\, y_b\nonumber\\
16\pi^2\,\frac{dA_\tau}{d\ln\mu}&=&
3g_2^2(2M_2y_\tau-A_\tau)+\frac{9}{5}g_1^2(2M_1y_\tau-A_\tau)\nonumber\\
&&+2\gz^2(Q_{H_d}^2+Q_L^2+Q_{e^c}^2)(2\mzp y_\tau-A_\tau)\nonumber\\
&&+12 A_\tau y_\tau^2+3y_b^2 A_\tau+
6 A_b y_b\, y_{\tau}+\lambda^2 A_\tau+
2 A_\lambda\lambda\, y_\tau\nonumber\\
16\pi^2\,\frac{dA_{\lambda}}{d\ln\mu}&=&3g_2^2(2M_2\lambda-A_\lambda)+
\frac{3}{5}g_1^2(2M_1\lambda-A_\lambda)\nonumber\\
&&+2\gz^2(Q_{S}^2+Q_{H_u}^2+Q_{H_d}^2)(2\mzp \lambda-A_\lambda)\nonumber\\
&&+A_\lambda(12\lambda^2+3y_t^2+3y_b^2+y_\tau^2+n_Ey_E^2+3 n_D y_D^2)\nonumber\\
&&+\lambda(6 y_tA_t+6 y_bA_b+2 y_\tau A_\tau+2 n_E y_E A_E+6n_Dy_DA_D)\nonumber\\
16\pi^2\,\frac{dA_{y_D}}{d\ln\mu}&=&\frac{16}3g_3^2(2M_3y_D-A_D)+\frac{6}{5}
\left(Y_D^2+Y^2_{D^c}\right)g_1^2(2M_1y_D-A_D)\nonumber\\
&&+2\gz^2(Q_{S}^2+Q_D^2+Q_{D^c}^2)(2\mzp y_D-A_D)+A_D(2\lambda^2+n_Ey_E^2+(9n_D+6) y_D^2)\nonumber\\
&&+y_D(4\lambda A_\lambda+2 n_E y_E A_E)\nonumber\\
16\pi^2\,\frac{dA_{y_E}}{d\ln\mu}&=&\frac{6}{5}\left(Y_E^2+Y^2_{E^c}\right)g_1^2(2M_1y_E-A_E)+
2\gz^2(Q_{S}^2+Q_E^2+Q_{E^c}^2)(2\mzp y_E-A_E)\nonumber\\
&&+A_E(2\lambda^2+3n_Dy_D^2+(3n_E+6) y_E^2)+y_E(4\lambda A_\lambda+6 n_D y_D A_D).
\end{eqnarray}

\end{appendix}

\end{document}